\documentclass[preprint,12pt,authoryear]{elsarticle}
\usepackage{amssymb}
\usepackage{amsmath}
\usepackage{aas_macros}
\usepackage{color}
\usepackage{subcaption}

\usepackage{lineno}
\usepackage{graphicx}
\usepackage{txfonts}
\usepackage{multirow}

\begin{document}

\defcitealias{Marschall2016}{M2016}
\defcitealias{Marschall2017}{M2017}

\begin{frontmatter}
\title{Limitations in the determination of surface emission distributions on comets through modelling of observational data - A case study based on Rosetta observations.}

\author[1,2]{Raphael Marschall}
\author[3]{Ying Liao}
\author[4]{Nicolas Thomas}
\author[5]{Jong-Shinn Wu}

\address[1]{Southwest Research Institute, 1050 Walnut St, Suite 300, Boulder, CO 80302 , marschall@boulder.swri.edu}
\address[2]{International Space Science Institute, Hallerstrasse 6, CH-3007 Bern, Switzerland}
\address[3]{Graduate Institute of Astronomy, National Central University, Taiwan}
\address[4]{Physikalisches Institut, Sidlerstr. 5, University of Bern, CH-3012 Bern, Switzerland}
\address[5]{Department of Mechanical Engineering, National Chiao Tung University, Hsinchu, Taiwan}

%\date{Received XXXX; accepted XXXX}
% \abstract{}{}{}{}{} 
% 5 {} token are mandatory

\begin{abstract}
{The European Space Agency's (ESA) Rosetta mission has returned a vast data set of measurements of the inner gas coma of comet 67/Churyumov–Gerasimenko. These measurements have been used by different groups to determine the distribution of the gas sources at the nucleus surface. The solutions that have been found differ from each other substantially and illustrate the degeneracy of this issue.}\\
   {It is the aim of this work to explore the limitations that current gas models have in linking the coma measurements to the surface. In particular, we discuss the sensitivity of Rosetta's ROSINA/COPS, VIRTIS, and MIRO instruments to differentiate between vastly different spatial distributions of the gas emission from the surface. \\}
   {We have applied a state of the art 3D DSMC gas dynamics code to simulate the inner gas coma of different models that vary in the fraction of the surface that contains ice and in different sizes of active patches. These different distributions result in jet interactions that differ in their dynamical behaviour. We have then produced synthetic measurements of Rosetta's gas instruments. By comparing the different models we probe the limitations of the different instruments to variations in the emission distribution.\\}
   {We have found that ROSINA/COPS measurements by themselves cannot detect the differences in our models. While ROSINA/COPS measurements are important to constrain the regional inhomogeneities of the gas emission, they can by themselves not determine the surface-emission distribution of the gas sources to a spatial accuracy of better than a few hundred metres ($\simeq 400$~m $\sim 50$~MFP). Any solutions fitting the ROSINA/COPS measurements is hence fundamentally degenerate, be it through a forward or inverse model. Only other instruments with complementary measurements can potentially lift this degeneracy as we show here for VIRTIS and MIRO. In particular, we find that MIRO is the only instrument that can distinguish between most of our models. Finally, as a by-product, we have explored the effect of our activity distributions on lateral flow at the surface that may be responsible for some of the observed aeolian features.}
\end{abstract}


\end{frontmatter}

%
%-------------------------------------------------------------------
%%%%%%%%%%%%%%%%%%%%%%%%%%%%%%%%%%%%%%%%%%%%%%%%%%%%%%%%%%%%%%%%%%%%%%
%%%%%%%%%%%%%%%%%%%%%%%%%%%%%%%%%%%%%%%%%%%%%%%%%%%%%%%%%%%%%%%%%%%%%%
%%%%%%%%%%%%%%%%%%%%         INTRODUCTION        %%%%%%%%%%%%%%%%%%%%%
%%%%%%%%%%%%%%%%%%%%%%%%%%%%%%%%%%%%%%%%%%%%%%%%%%%%%%%%%%%%%%%%%%%%%%
%%%%%%%%%%%%%%%%%%%%%%%%%%%%%%%%%%%%%%%%%%%%%%%%%%%%%%%%%%%%%%%%%%%%%%
\section{Introduction}\label{introduction}
The European Space Agency's (ESA) Rosetta mission escorted comet 67P/Churyumov–Gerasimenko (hereafter 67P) on its orbit around the Sun for two years. The mission provided an unprecedented amount of data on, among others, the comet's neutral gas coma. By observing the comet in close proximity covering heliocentric distances from $\sim3.5$AU inbound, through to perihelion at $1.24$~AU, to $\sim3.5$~AU outbound we are able to study the development, change and fading of the gas coma. The vast amount of data combined with a high definition shape model of the nucleus surface \citep{Preusker2017} spurs hope of linking the in-situ and remote sensing gas measurements to the surface and determining the geographical location of the emission and in particular variations therein. Models have played a key role in this attempt to link coma measurements to the surface.

Simulating and studying the flow of gas from the cometary surface into $3$D space is complex and depends on the scale at which the coma is to be studied. When ice sublimates into the vacuum it forms a non-equilibrium boundary layer, the "Knudsen layer", with a scale height of $\sim 10$ to $100$ mean free paths (MFP, $\lambda$) \citep{Ytrehus1975,Davidsson2008}. Within this layer, the velocity distribution function (VDF) is strongly non-Maxwellian. This layer can become infinitely thick when the production rate is low. Gas flows can cover the fully collision-less free flow regime, in which $\lambda \sim \infty$, to the fluid regime with a very short MFP. To determine the flow regime of a specific case the Knudsen number, $Kn$, is defined as
\begin{equation}\label{eq:knudsen-number}
  Kn = \frac{\lambda}{L} \quad ,
\end{equation}
where $L$ is the characteristic size of the studied system, in the case of comet 67P case a few kilometres to tens of kilometres. Irrespective of the flow regime, the Boltzmann equations need to be satisfied but for certain regimes, approximations can be made to calculate the flow field more easily. When the Knudsen number is small ($Kn < 0.1$) the gas flow is in the fluid or continuum regime and the flow can be accounted for by solving the Euler equations or Navier-Stokes equations. Such (multi-)fluid models for comets with high production rates have been studied by e.g. \citet{Marconi1982,Marconi1983,Kitamura1986,Kitamura1987,Crifo1997a,Gombosi1985,Crifo1997b} for the case of dusty-gas flows and by e.g. \citet{Combi1996,Tenishev2008,Shou2016} for the case of multi-species gas flows. On the other hand, when the Knudsen number is very large ($Kn \gg 10$) collisions can be neglected and we consider this the free molecular flow regime. Historically, these were the first models studied e.g. by \citet{Eddington1910,Haser1957,WallaceMiller1958} as they are the simplest models that can be considered. For the intermediate regime where $0.1<Kn<10$ the Euler and Navier-Stokes equations are no longer valid and the flow can also not be considered to be in the free molecular flow regime yet. Thus in this intermediate regime, the kinetics of molecule-molecule collisions need to be accounted for.

 First models using a statistical Monte Carlo approach were studied by \citet{CombiDelsemme1980,Kitamura1985, CombiSmyth1988, Hodges1990} (test particle approach), \citet{Xie1Mumma1996a,Xie1Mumma1996b} (direct simulation Monte Carlo; DSMC). Comparisons of DSMC with fluid models were performed e.g. by \citet{Crifo2002}. The state of the art method for studying non-equilibrium though is Direct Simulation Monte Carlo (DSMC) first proposed by Bird \citep[see][]{Bird1976} and which has since been applied by many authors \citep{Combi1996,SkorovRickman1999,Crifo2002,Crifo2003,Tenishev2008,Tenishev2011,Davidsson2008,Zakharov2008,Fougere2013} especially as the computational power available has significantly increased in the past decades.\\

\begin{figure}
  \includegraphics[width=\linewidth]{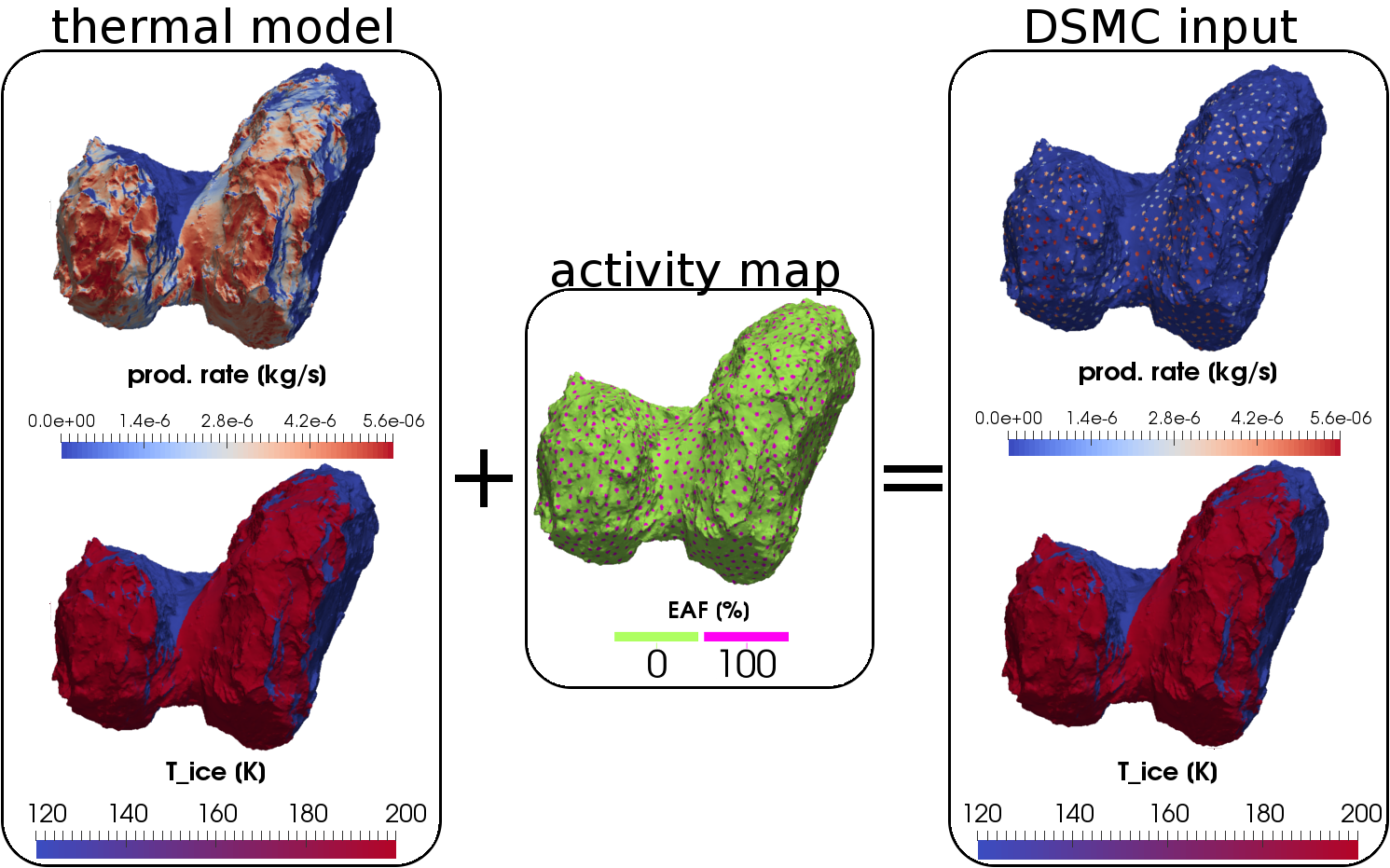}
  \caption{Illustration of how an activity map is convolved with the result of a thermal model to give the scaled production rates that go into a gas dynamics model. Notably, the production rates are scaled with the activity map while the temperature of the sublimating surface is left unchanged.}
  \label{fig:process} 
\end{figure}

Concerning 67P three instruments are of particular relevance for studying the gas coma. First, there is the Rosetta Orbiter Spectrometer for Ion and Neutral Analysis (ROSINA). ROSINA/COPS consists of the Double Focusing Mass Spectrometer (DFMS), the Reflectron-Type Time-Of-Flight (RTOF) mass spectrometer, and the COmet Pressure Sensor (COPS) to measure the composition and total density of volatiles from in-situ, i.e. at the location of the spacecraft \citep{Balsiger07}. For models, in particular, COPS and DFMS are of importance because together they give the local gas number density at the spacecraft location for different species, including the major molecules, H$_{2}$O, CO$_{2}$, CO \citep{Haessig15}, and O$_2$ \citep{Bieler2015Natur} as well as many minor species \citep[e.g.][]{LeRoy2015}. Second, there is the Visible and Infrared Thermal Imaging Spectrometer (VIRTIS) \citep{Coradini2007}. VIRTIS-M-IR is of particular interest because it acquired spectrally and spatially resolved images (i.e. cubes) of the coma. As shown in \citet{Migliorini2016} the spectra can be converted into H$_{2}$O and CO$_{2}$ gas column densities which in combination with the spatial resolution of VIRTIS-M, results in H$_{2}$O and CO$_{2}$ maps of the coma. Third, there is the Microwave Instrument on the Rosetta Orbiter (MIRO) which consists of two heterodyne receivers: a sub-millimetre spectrometer operating at a centre band frequency of 562~GHz (0.5 mm); and a millimetre spectrometer at 188~GHz (1.6~mm) \citep{Gulkis2007}. The MIRO spectra can be used to produce H$_2$O column densities \citep{Biver2015} and even along the line-of-sight (LOS) gas properties (number density, temperature, and speed along the LOS) as detailed in \citet{marschall2019}. Here we will concentrate on H$_2$O only but our conclusions apply in general as they are in no way specific to water. \\

Different authors \citep{Bieler2015,Marschall2016,Fougere2016,Marschall2017,Zakharov2018} have used similar approaches to fit Rosetta gas data (primarily ROSINA/COPS\&DFMS) and by doing so constrained the emission distribution on the nucleus surface. How this is generally done is illustrated in Fig.~\ref{fig:process}. The illumination conditions of the nucleus (taking into account self-shadowing) are used to calculate the energy input to the surface which is fed into a thermal model that gives the local gas production rates and surface temperatures assuming, in this case, a pure ice surface. The left column in Fig.~\ref{fig:process} shows the resulting production rates and temperature of the sublimating surface. Notably, only the gas production rates are scaled, while the temperatures of the sublimating surface are kept unchanged. To scale the model to fit the data a free parameter needs to be introduced. We call this the effective active fraction (EAF). The EAF can be thought of as the areal fraction (akin to a checkerboard pattern) of a surface facet that is pure ice and can in principle be set individually for each facet. These free parameters are called "relative activity" in \citet{Fougere2016} and "inhomogeneity factors" in \citet{Zakharov2018} but are the same concepts as ours. Fundamentally the introduction of these parameters allows one to separate effects of the thermal model itself with those arising due to e.g. a different ice content of different areas or a variably thick dust cover quenching the activity in different places. The sum of all of these free parameters constitutes what can be called an activity map. An example of such a map can be seen in the centre column of Fig.~\ref{fig:process}. In principle, this map should not change in time, at least on short time scales (rotation period of the comet). When convolving such a map with the results from the thermal model we reach the actual production rates at the surface that go into the $3$D gas dynamics models that determine the flow field and the predictions which can be compared to the Rosetta measurements. When comparing the resulting activity maps (\citet[Fig. 4]{Marschall2016}, \citet[Fig. 5]{Fougere2016}, \citet[Fig. A.6]{Zakharov2018}) it becomes clear that there does not seem to be a unique solution. Qualitatively the maps of \citet{Marschall2016} and \citet{Fougere2016} are similar but very different to the one by \citet{Zakharov2018}. A further map of \citet[Fig. 5]{Laeuter2019} who have used an analytical approach assuming collision-less outflow of the gas to invert ROSINA/COPS data for the determination of the surface sources is not directly comparable to the above-mentioned maps because they are in flux rather than EAF. Further discussion on this last model can be found in Appendix~\ref{sec:LSFM}. But the maps from \citet{Laeuter2019} can be compared to the erosion maps presented in \citet[Fig. 9]{Combi2020}, which is an improvement and extension of the work by \citet{Fougere2016}. Though the flux/erosion maps of \citet{Laeuter2019} and \citet{Combi2020} agree in broad terms as the latitudes at which the mass loss occurs there remain significant differences as to the exact location of the erosion. These differences result in different implications for the surface morphology and observed surface activity. In any case, the published emission distributions demonstrate a need to better understand the limitations of these models and possible ways of lifting the inherent degeneracies of the solutions.\\

This work focuses exclusively on the volatile component of cometary activity and the associated Rosetta gas instruments. It is clear that as dust is dragged in the gas flow it can provide further constraints on the locations where mass loss occurs. Implications of our work on dust observations - in particular, OSIRIS - will be explored in future work.\\

\begin{figure}
  \includegraphics[width=\linewidth]{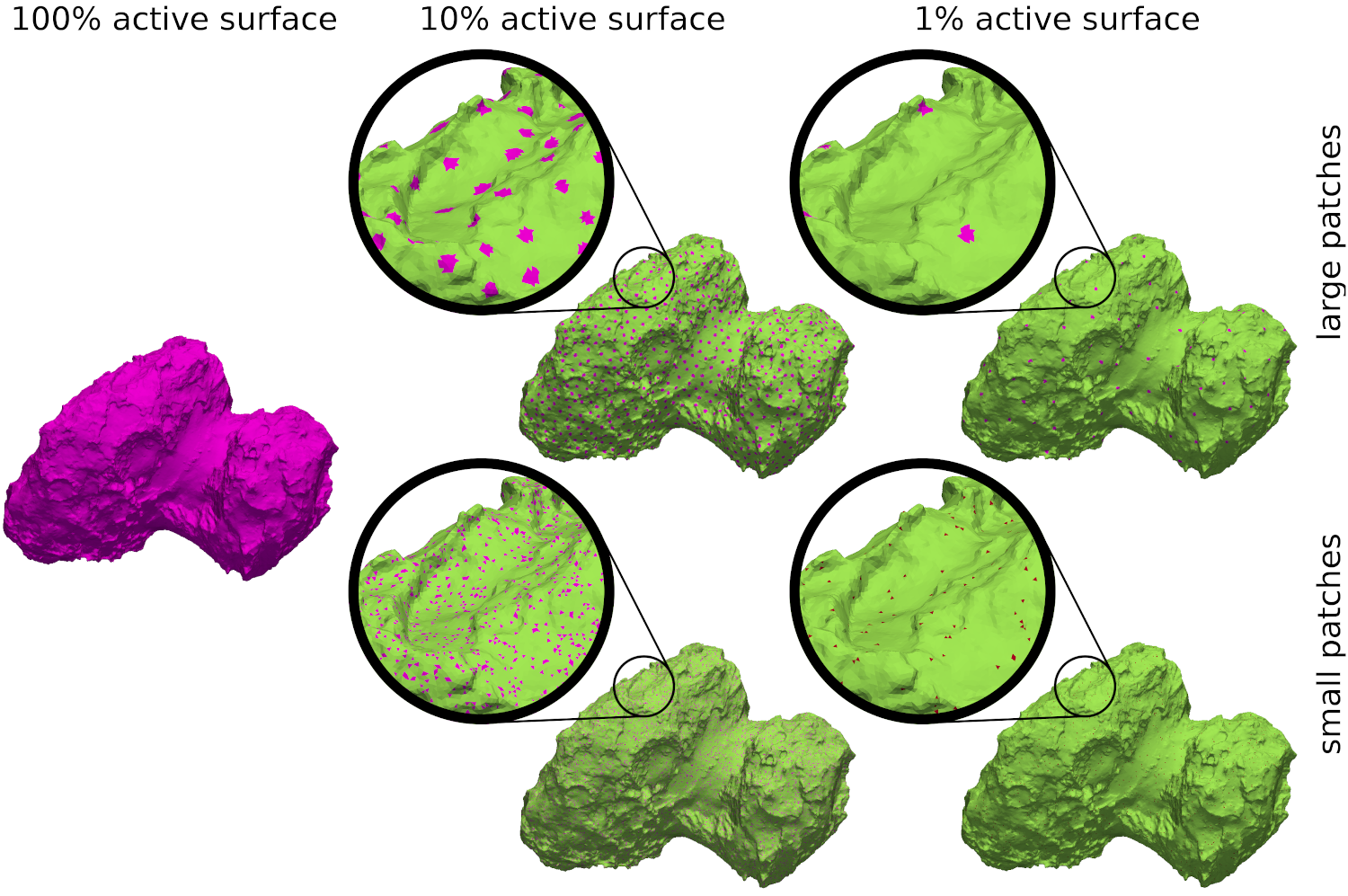}
  \caption{Five basic activity maps used in this map. Purple areas have the potential for outgassing with effective active fractions (EAF) that are non-zero, while green areas are fully inert, thus having an EAF of zero.}
  \label{fig:eaf-all} 
\end{figure}

In this work, we will explore the limits of resolving inhomogeneities of the emission distribution at the surface. We will show that although ROSINA/COPS can detect regional/large scale inhomogeneities, it is rather insensitive to inhomogeneity on scales smaller than a few hundred meters and that this is a physical limit rather than the result of the numerics of models. This implies that neither forward nor inverse models can provide unique solutions with high spatial resolution when incorporating ROSINA/COPS data only. Furthermore, we will explore how including VIRTIS-M-IR and MIRO data in a multi-instrument approach can lift some of the degeneracies. And finally, we will discuss the implications of inhomogeneous outgassing patterns on lateral flow across the surface.

%%%%%%%%%%%%%%%%%%%%%%%%%%%%%%%%%%%%%%%%%%%%%%%%%%%%%%%%%%%%%%%%%%%%%%
%%%%%%%%%%%%%%%%%%%%%%%%%%%%%%%%%%%%%%%%%%%%%%%%%%%%%%%%%%%%%%%%%%%%%%
%%%%%%%%%%%%%%%%%%%%             MODELS          %%%%%%%%%%%%%%%%%%%%%
%%%%%%%%%%%%%%%%%%%%%%%%%%%%%%%%%%%%%%%%%%%%%%%%%%%%%%%%%%%%%%%%%%%%%%
%%%%%%%%%%%%%%%%%%%%%%%%%%%%%%%%%%%%%%%%%%%%%%%%%%%%%%%%%%%%%%%%%%%%%%
\section{Method and model setups} \label{sec:method-and-model-setup}
The first fundamental question we seek to answer is whether a unique surface distribution can be inferred from the ROSINA/COPS data alone. The possibility to establish uniqueness is also important to the question of whether or not this problem can be inverted. It is more straight forward though to show non-uniqueness by finding different surface distributions that result in the same measurements.
We, therefore, seek initial states that are clearly different from each other but result in the same predicted measurements within a given error. We achieve strongly different initial states (as described in more detail later) by considering activity maps that have large portions (up to $99\%$) of the surface being inactive compared to other maps. Also, we examine cases that allow us to infer the resolution limit. We can thus be confident should we find the same predicted measurements that we have indeed shown that the data cannot lead to unique solutions below a certain resolution.\\

The production rate at 67P spans a large range from low (several kg s$^{-1}$) to high (hundreds kg s$^{-1}$) activity and therefore the gas flow in principle transitions from either fluid or collisional to a free adiabatic outflow. Thus our preferred method is DSMC because it intrinsically covers all of these regimes. For high production rates a fluid approach might be computationally more economical though. The code we are using, named ultraSPARTS\footnote{by Plasma T.I., http://plasmati.com.tw} a further development of PDSC$^{++}$ \citep{Su13}, is based on the PDSC code developed by Wu and co-workers \citep{Wu03,Wu04,Wu05}. We have used a decimated shape model of SHAP7 \citet{Preusker2017}. The full resolution SHAP7 model has about 44 million facets which is far larger than the computational capacities we currently have. We thus use a decimated model with $\sim 440,000$ facets. The resulting 3D unstructured simulation grid that is tied to the shape model and goes out to 10~km from the nucleus centre has over 13 million cells. For more details on our general approach and gas dynamics code we refer to \citet{Marschall2016} and the references therein.\\

What all models by the different groups have in common is that on a global average the EAF is only a few percents (typically $1\%$). This implies that the comet produces the amount of gas that is equivalent to $1\%$ of the surface being a pure ice surface and the rest being inert. This naturally raises the question of how this low percentage is distributed on the surface and how we would detect this. We have thus devised a variety of extreme test cases to examine this exact question. Figure~\ref{fig:eaf-all} illustrates the five basic distributions of ice we have tested. We have varied three basic parameters. First, the percentage of the surface that has the potential for activity. The one extreme here is that the entire surface has $1\%$ of ice coverage that is not resolved by our shape model (left column in Fig.~\ref{fig:eaf-all}). The other extreme is that $1\%$ of the surface is pure ice with the remaining $99\%$ of the surface being totally inert and hence having an EAF of zero (right column in Fig.~\ref{fig:eaf-all}). We have also looked at the intermediate case where $10\%$ of the surface is active (centre column in Fig.~\ref{fig:eaf-all}). First simulations using such distributions were run by \citet{Liao2018}. The second variable we have explored is how the ice is agglomerated on the surface. In the one extreme, we have used the smallest unit (the surface facets) acting as pure ice patches (bottom row in Fig.~\ref{fig:eaf-all}). In the other extreme, we have uniformly spaced larger patches (consisting of multiple facets) over the surface (top row in Fig.~\ref{fig:eaf-all}). For the $10\%$ active surface case we have used $2000$ large patches. In the $1\%$ active surface case we have used $200$ large patches were used. The resulting five activity maps have been scales such that they globally produce the same amount of gas as the comet rotates. This was also the limiting factor in the number of large patches. We assure that the global production rate of all cases as the comet rotates are as similar as possible such that our results hold for a full comet rotation. Here we are interested in activity maps that are radically different but are potentially indistinguishable. The detailed characteristics of the patches - such as their size - are summarised in Tab.~\ref{tab:patch-properties}. The third variable we varied is the global production rate. We have performed two sets of simulations for each of the activity maps with a low global production rate of $2$~kg~s$^{-1}$ (corresponding to approximately $3.0$~AU inbound [$\sim$ November 16, 2014] and $2.8$~AU outbound [$\sim$ April 9, 2016]) and a high global production rate $200$~kg~s$^{-1}$ (corresponding to approximately $1.35$~AU inbound [$\sim$ July 14, 2015] and $1.5$~AU outbound [$\sim$ October 22, 2015] estimated from the values given \cite{Hansen2016}). We are therefore confident that the results we present here apply to almost the entire orbit. The surface temperatures are calculated from the energy balance for each of the above mentioned inbound heliocentric distances. Shadowed surface areas are assumed to have a surface temperature of $100$~K. For the models with inactive areas the temperatures were assigned as if they were active but no flux was assigned. For all results presented in this work the position of the Sun relative the comet has been kept fixed at a sub-solar latitude of $30^\circ$ and sub-solar longitude of $225^\circ$ in the Cheops fixed frame \citep{Preusker2015}.\\

\begin{table}[]
\begin{center}
\caption{Patch properties of the different models.}\label{tab:patch-properties}
\begin{tabular}{l|r|r|r|r}
patch type             & ASA$^{(a)}$ & D$^{(b)}$ [m] & d$^{(c)}$ [m]           & A$^{(d)}$ [m$^2$] \\ \hline \hline
\multirow{2}{*}{large} & 1\%         & 500       & \multirow{2}{*}{57} & \multirow{2}{*}{2,550}        \\
                       & 10\%        & 160       &                     &                              \\\hline
\multirow{2}{*}{small} & 1\%         & 107       & \multirow{2}{*}{12}  & \multirow{2}{*}{115}         \\
                       & 10\%        & 33        &                     &                             
\end{tabular}
\end{center}
{\small \textbf{Notes.} $^{(a)}$ active surface area, $^{(b)}$ mean minimum separation between two patches, $^{(c)}$ mean equivalent patch diameter assuming a circular patch, $^{(d)}$ mean patch area.}
\end{table}

\section{General theoretical considerations of interacting jets}\label{sec:theory-jets}
\begin{figure}
  \includegraphics[width=\linewidth]{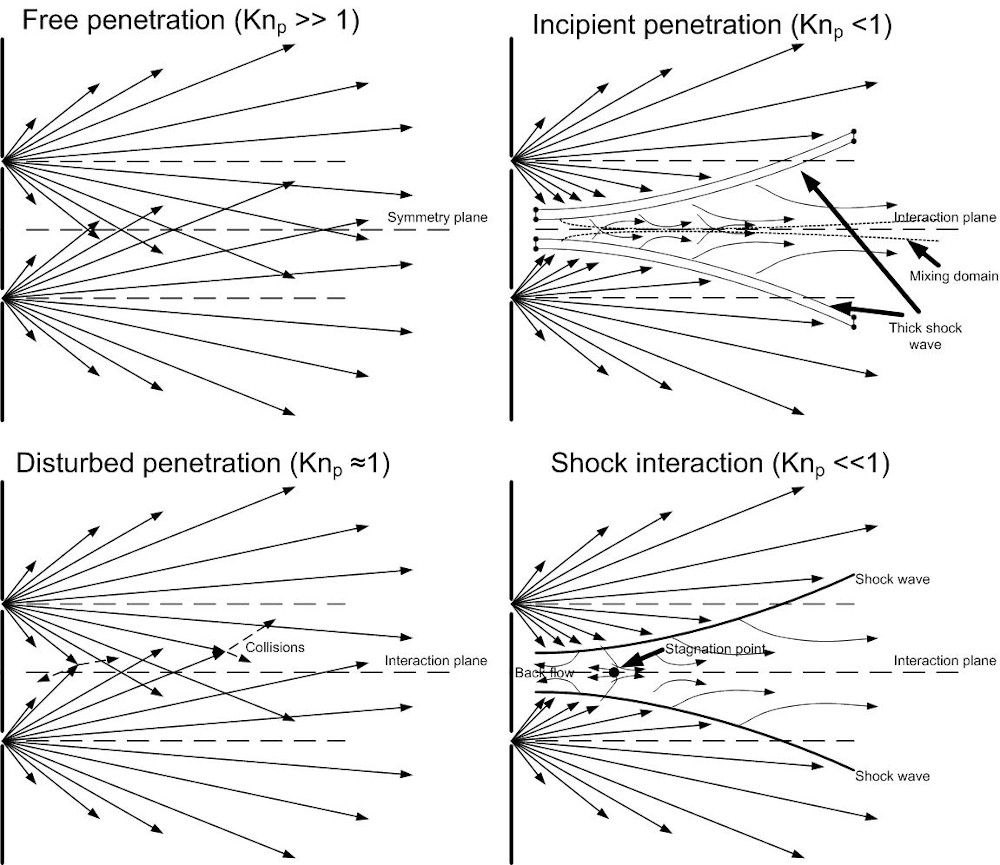}
  \caption{Illustration of the four different jet interaction regimes as defined by \cite{DankertKoppenwallner1984} from free penetration regime with $Kn_p >> 1$ to the shock interaction regime with $Kn_p << 1$. This illustration has been adapted from \cite{DankertKoppenwallner1984} and \cite{Dettleff1991}.}
  \label{fig:jet-interactions} 
\end{figure}

The models we have set up essentially represent different types of jet-like emission patterns. These jets interact with each other depending on the type of flow regime we are in. To gauge what we can expect from these different models we will first explore what is already known about jet interactions. The interaction of jets or plumes of gas has been studied for at least 35 years mainly in the context of the interaction of thrusters on space probes. \cite{DankertKoppenwallner1984} distinguish between four different regimes for the interaction of two jets as illustrated in Fig.~\ref{fig:jet-interactions}. To differentiate between the different regimes the penetration Knudsen number, $Kn_p$, is defined. It can be understood similarly to the Knudsen number defined in Eq.~\ref{eq:knudsen-number} and is the ratio of the mean free path (MFP) at the surface of the particles from the one source to the distance from the interaction plane to the centre line of the next nearest source (half of the distance between the two sources). This assumes that the two sources are equally strong and that the surface is not significantly bent (i.e. large nucleus radius). This is, of course, a simplification but allows the determination of different interaction regimes. The four regimes can be characterised as follows:
\begin{itemize}
    \item \textbf{Free penetration:} For very large values of $Kn_p$ (top left in Fig.~\ref{fig:jet-interactions}) - corresponding to low gas production rates - both jets are in the free molecular flow regime and penetrate each other without any resistance. The total gas densities, in this case, are simply the sum of the densities of each jet individually. 
    \item \textbf{Disturbed penetration:} When the emission increases to the point where $Kn_p$ is of the order of unity (bottom left in Fig.~\ref{fig:jet-interactions}) the two jets start interacting. Molecules start colliding with each other and the flows disrupt each other close to the interaction plane. In this case, the total gas densities are no longer the sum of the densities of the individual unperturbed jets.
    \item \textbf{Incipient penetration:} Further increasing the gas production rate of the two jets and thus further lowering $Kn_p$ to less than unity (top right in Fig.~\ref{fig:jet-interactions}) results in the formation of a mixing domain close to the interaction plane and a thick shock wave region that separates the mixing domain from the jet regions.
    \item \textbf{Shock interaction:} Finally, when the gas emission is further increased such that $Kn_p << 1$ (bottom right in Fig.~\ref{fig:jet-interactions}) a strong shock wave forms separating the two jets. Within the shock there exist a point where the flow is stagnant. Below that the flow is redirected back to the surface. If there are more than two interacting jets this structure also becomes more complex.
\end{itemize}

\begin{table}[]
\caption{Lower limit for the penetration Knudsen number, $Kn_p$, for the different models}\label{tab:penetration-knudsen-number}
\begin{tabular}{llr|rr}
activity              & patch                  & ASA$^{(a)}$  & $Kn_p >$ & D$^{(b)}$ [MFP]\\ \hline\hline
\multirow{4}{*}{high} & \multirow{2}{*}{large} & $1\%$  & $0.04$        &  $50$ \\
                      &                        & $10\%$ & $0.1$         &  $20$ \\ 
                      & \multirow{2}{*}{small} & $1\%$  & $0.2$         &  $10$ \\
                      &                        & $10\%$ & $0.4$         &  $5$ \\\hline
\multirow{4}{*}{low}  & \multirow{2}{*}{large} & $1\%$  & $3$           &  $0.67$ \\
                      &                        & $10\%$ & $6$           &  $0.33$ \\
                      & \multirow{2}{*}{small} & $1\%$  & $15$          &  $0.13$ \\
                      &                        & $10\%$ & $30$          &  $0.067$
\end{tabular}\\
{\small \textbf{Notes.} $^{(a)}$ active surface area, $^{(b)}$ mean minimum separation between two patches divided by the mean free path (MFP).}
\end{table}

\citet{Dettleff1991} has outlined in Eq.~(91)-(93) how the $Kn_p$ can be calculated. As mentioned above reality does not satisfy some of the simplifications assumed by \citet{Dettleff1991}. To make a robust as possible estimate of $Kn_p$ we have therefore chosen to measure the MFP at the surface ($\theta = \pi/2$ in \citet{Dettleff1991}) of the source rather than at the interaction plane. This circumvents the interpretation as to where the interaction plane is which is complex if the sources are not equally strong. To mitigate the effects of the simplified assumptions on the definition of $Kn_p$ we have measured $Kn_p$ for the patch closest to the sub-solar point (which in our cases is in the Imhotep region). This minimises the relative strength difference between neighbouring sources and Imhotep comes close to being a flat surface. We estimate $Kn_p$ for the different activity levels, patch sizes, and active surface areas. Our estimates presented in Tab.~\ref{tab:penetration-knudsen-number} represent a lower limit for $Kn_p$ because they are measured at the patch with the highest production rate. We can observe that in the low activity models we find ourselves in the regime between free and disturbed penetration while in the high activity models we find ourselves between the regimes of incipient penetration and shock interaction. In the latter cases, we can expect shocks to form. We can hence expect to observe clear differences between the two activity levels. In both activity cases, $Kn_p$ decreases as the active surface area decreases and the patch size increases. It is also notable that only the most extreme cases with activity coming from only $1\%$ of the surface in large patches do we come close to a new regime different from the other cases. Because the other three models within each activity level group are clearly in one regime we should expect these models to be hard to differentiate - in particular in the low activity case. In the high activity case the fact that we can expect shocks - be it mild ones - should result in increasing contrast between jet regions as $Kn_p$ decreases to the point where we have almost reached the shock interaction regime. The table further shows the mean distance between patches in MFP.

%%%%%%%%%%%%%%%%%%%%%%%%%%%%%%%%%%%%%%%%%%%%%%%%%%%%%%%%%%%%%%%%%%%%%%
%%%%%%%%%%%%%%%%%%%%%%%%%%%%%%%%%%%%%%%%%%%%%%%%%%%%%%%%%%%%%%%%%%%%%%
%%%%%%%%%%%%%%%%%%%%             Results         %%%%%%%%%%%%%%%%%%%%%
%%%%%%%%%%%%%%%%%%%%%%%%%%%%%%%%%%%%%%%%%%%%%%%%%%%%%%%%%%%%%%%%%%%%%%
%%%%%%%%%%%%%%%%%%%%%%%%%%%%%%%%%%%%%%%%%%%%%%%%%%%%%%%%%%%%%%%%%%%%%%
\section{Results}\label{sec:results}
\subsection{The ROSINA/COPS view} \label{sec:ROSINA-results}
\begin{figure}
  \includegraphics[width=\linewidth]{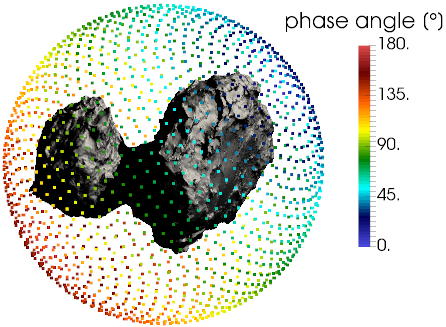}
  \caption{Illustration of the points along the artificial orbit from which gas number density measurements were take to simulate what ROSINA/COPS would measure.}
  \label{fig:orbit} 
\end{figure}
Different authors \citep[e.g.][]{Marschall2016,Fougere2016,Zakharov2018} have used ROSINA/COPS \& DFMS measurements of the local gas density to constrain the emission distribution at the surface. We shall, therefore, examine here to what extent this is even possible. Indeed we will see that to a large extent local gas density measurements are insensitive to the kind of local inhomogeneities we examine here but can detect larger scale/regional variation well (see Sec.~\ref{sec:regional-inhom}). To compare the predicted measurements from the different models we have defined an artificial orbit that samples the gas coma at $10$~km as a function of phase angle. The sampling points are illustrated in Fig.~\ref{fig:orbit}. For phase angle steps every $5^\circ$ synthetic measurements were taken with a $5^\circ$ spacing along constant phase angle. We will focus on analysing the results up to phase angles of $90^\circ$ because we want to examine the dayside activity and the corresponding emission pattern at the surface. This synthetic orbit represents in a sense the most optimistic case for sampling the cometary gas coma as it covers the highest activity region from low phase angles to the terminator close to the surface. If differences are not seen in this kind of orbit it would not be seen either in an orbit farther away or confined to a smaller phase angle range. Hence the actual orbit flown by Rosetta represents less favourable conditions than the one presented here.
\begin{figure}
  \includegraphics[width=\linewidth,trim={0 0 0 2.3cm},clip]{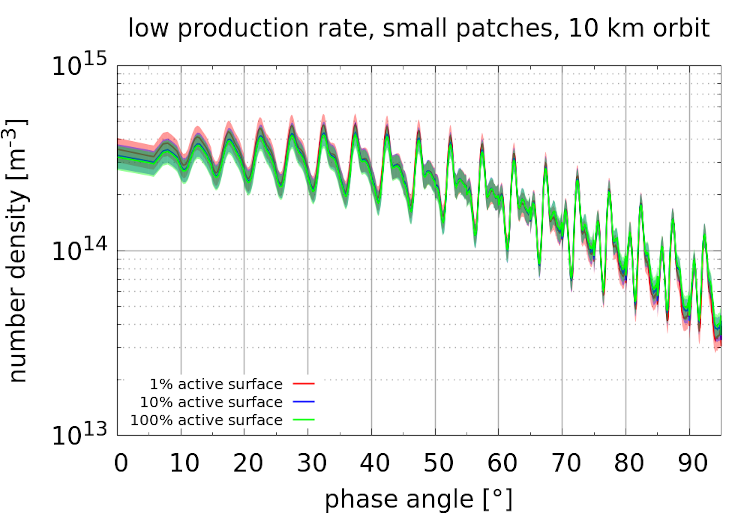}
  \caption{Local gas number density [m$^{-3}$] as a function of phase angle comparing the 100\% (green), 10\% (blue), and 1\% (red)  active surface models with small patches for the low activity case. The bands indicate $\pm 15\%$ error intervals. Measurements are taken every $5^\circ$ in phase angle. The values between each $5^\circ$ interval correspond to the measurements along the circle of constant phase angle (i.e. the values of the interval [$5^\circ$,$10^\circ$) correspond to the measurements taken along the $5^\circ$ circle)}
  \label{fig:rosina-lowQ_SP} 
\end{figure}

\begin{figure}
  \includegraphics[width=\linewidth,trim={0 0 0 2.3cm},clip]{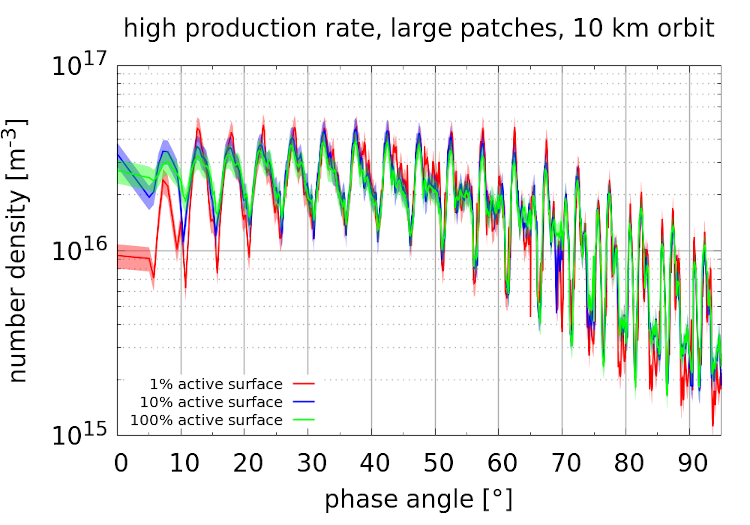}
  \caption{Local gas number density [m$^{-3}$] as a function of phase angle comparing the 100\% (green), 10\% (blue), and 1\% (red)  active surface models with large patches for the high activity case. The bands indicate $\pm 15\%$ error intervals.}
  \label{fig:rosina-highQ_LP} 
\end{figure}

Figures~\ref{fig:rosina-lowQ_SP} and \ref{fig:rosina-highQ_LP} show the local gas number density as a function of the phase angle for the comparison of the 100\%, 10\%, and 1\%  active surface models with small patches for the low activity case and large patches for the high activity case. As explained above we take measurements every $5^\circ$ in phase angle. The values between each $5^\circ$ interval correspond to the measurements along the circle of constant phase angle (i.e. the values of the interval [$5^\circ$,$10^\circ$) correspond to the measurements taken along the $5^\circ$ circle). \\
These two figures show the extreme ends of the four combinations of patch size and activity level and thus show the range of the results. The low activity cases with small patches (Fig.~\ref{fig:rosina-lowQ_SP}) result in almost identical local number densities for the entire orbit respectively of the fraction of the surface that is active. This is caused by the fact that the large MFP of the gas molecules allows any inhomogeneities to be smoothed out within the first few kilometres above the surface. It is therefore impossible to determine the surface-emission distribution, in this case, using ROSINA/COPS data only. Concerning the other end of the spectrum of parameter space the high activity case with large patches shows a slightly different picture. The 100\% and 10\% cases are again qualitatively very similar but we see several deviations for the 1\% case. In the latter case, the inhomogeneities are preserved from the surface out to the spacecraft to a large extent because of the small MFP and large separation of the sources. This is in line with what we have expected from the theory of interacting jets presented in Sec.~\ref{sec:theory-jets}.
The other two combinations of patch size and activity level are shown in the appendix Figs.~\ref{fig:rosina-lowQ_LP} and \ref{fig:rosina-highQ_SP}. \\

\begin{figure}
  \includegraphics[width=\linewidth]{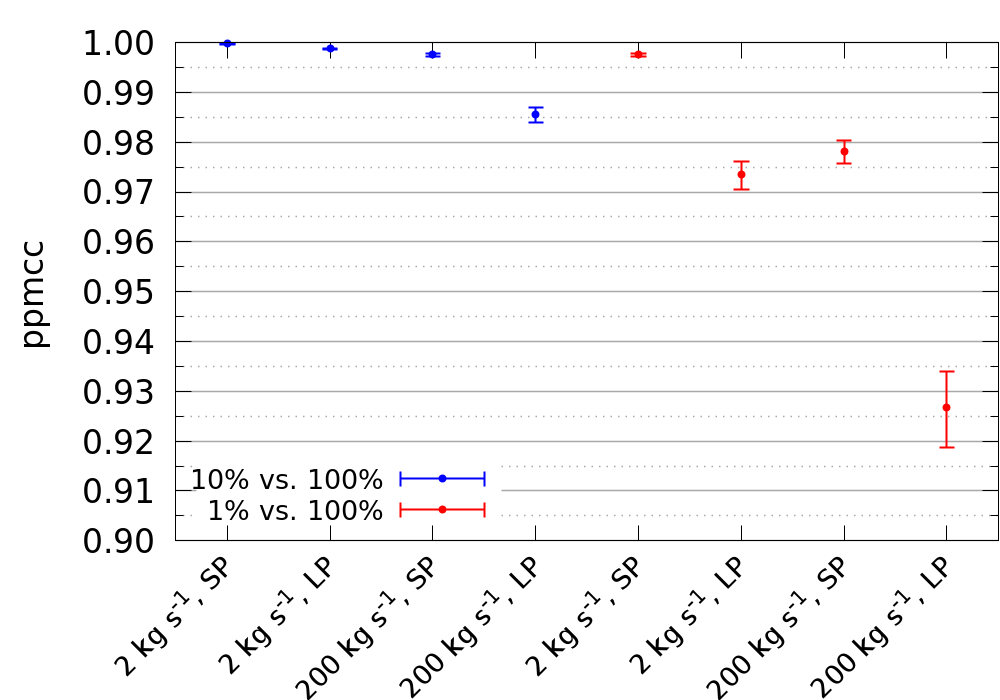}
  \caption{Pearson product-moment correlation coefficient (PPMCC) for the eight patchy cases comparing the 10\% (blue), and 1\% (red) active surface cases with small patches (SP) and large patches (LP) to the 100\% active surface case. The error bars represent a 2$\sigma$ confidence interval. The values have been calculated with the number densities at $10$~km and for phase angles covering $0-90^\circ$.}
  \label{fig:rosina-ppmcc} 
\end{figure}

For a more quantitative analysis of the goodness of the fit two methods have mainly been used in the literature. The first is the Pearson product-moment correlation coefficient (PPMCC) which measures whether two data sets (e.g. measurements and model predictions) are linearly correlated. The PPMCC takes the value of unity if the two data sets are perfectly correlated, zero if there is no correlation and $-1$ if they are perfectly anti-correlated. As an example \cite{Bieler2015} found a PPMCC of 0.93 for their best fit and \cite{Marschall2017} 0.96 for their best fit. We should mention here though, that the PPMCC depends heavily on the time span (number of data points) included for the correlation. The longer the time span the less diurnal variations influence the correlation. Therefore the PPMCC of different groups cannot be easily compared. But it can be used in a straightforward way to compare different models of a certain group that have the same data sampling and cover the same measurements. In Fig.~\ref{fig:rosina-ppmcc} we provide the PPMCC of the different models compared to the baseline model with $100\%$ active area. Clearly, all models provide very high correlations of above 0.92. All models with a $10\%$ active area correlate with values higher than $0.98$ and hence almost perfectly. The model with the lowest correlation by far is the one with $1\%$ active area and large patches for a high global gas production rate. 

\begin{figure}
  \includegraphics[width=\linewidth]{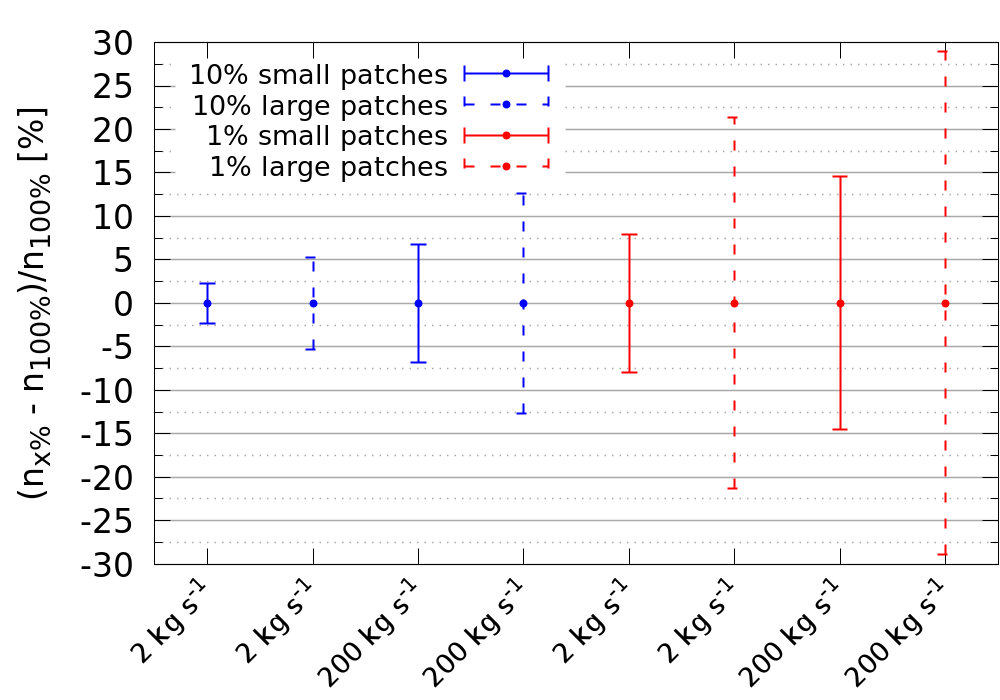}
  \caption{Mean relative difference between the number density of the eight patchy cases comparing the 10\% (blue), and 1\% (red) active surface cases with small patches (solid lines) and large patches (dashed lines) to the 100\% active surface case. The error bars represents the standard deviation of the number density differences.}
  \label{fig:rosina-delta} 
\end{figure}

The second measure often used is the mean relative difference between the model points and the measurements. For example \cite{Zakharov2018} find that their best fit models have mean relative differences of between $12-19\%$. Furthermore, according to \cite{Tzou2017} the error on ROSINA/COPS's nude gage is up to $15\%$ although the relative errors are probably smaller (M. Rubin, pers. comm.). Additionally, our DSMC models have errors on the order of $5-10\%$ arising from the accuracy of the collision model \citep{Finklenburg14}. From Fig.~\ref{fig:rosina-delta} we see that all but two models have mean deviations of less than $15\%$ and could therefore not be differentiated by measurements. Only the two extreme cases where only $1\%$ of the surface is active in large patches do we have larger deviations. These would be picked up in a rigorous analysis. We can, therefore, conclude that we cannot differentiate models with mean deviations of $<15\%$ and PPMCCs $>0.97$.

We have found strongly different activity maps that result in indistinguishable predicted measurements within the errors. Though we need to emphasise what we mean by "strongly different maps". There are different standards you may apply for assessing this. If your objective is to predict the local density measured at a spacecraft tens of kilometres from the surface and not make a strong link to the surface then you might not consider the presented maps different because, as we presented here, they do not differ in their predicted measurements. In this case, comparing the surface activity distribution (i.e. the size and distance of patches) to the MFP and the distance to the spacecraft is the relevant metric. If any inhomogeneities are on spatial scales small compared to the MFP and distance to the spacecraft then you could choose to consider them similar or the same. Contrary to that, here we seek a higher standard. Even if two maps result in the same predicted local gas densities at the spacecraft we can still conclude that the maps are strongly different if they imply different ground truths about the nucleus. In our case, the various maps are very different because depending on which one resembles the surface distribution closer will have a big influence on the implications we draw for the nucleus. A homogeneous nucleus, absent any evolutionary alterations, would result in a uniform activity distribution as e.g. our $100\%$ active area model. On the other hand, a heterogeneous nucleus would likely exhibit areas that are inert (or weakly active) and others that are highly active according to the ice content and thus resemble something closer to one of our $1\%$ active area models. Furthermore, these different maps will influence surface processes (such as lateral flow; see Sec.~\ref{sec:lateral-flow}) and thus in turn how we understand e.g. surface morphology. \\
All of this illustrates that based on ROSINA/COPS data alone it is not possible to differentiate between any of these models - arguably except for the most extreme cases where only $1\%$ of the surface is active in large patches. Extreme inhomogeneities of the surface-emission with a resolution lower than a few hundred metres can thus not be determined. This also implies that the ROSINA/COPS data simply cannot be inverted to give information about the emission distribution at the surface with a high spatial resolution. 
Any surface solution constrained only by ROSINA/COPS data thus remains fundamentally degenerate on spatial scales below a few hundred meters (tens of mean free paths), be it the result of an inverse model \citep[e.g.][]{Hansen2016,Kramer2017,Laeuter2019} or a forward model  \citep[e.g.][]{Fougere2016,Marschall2017,Zakharov2018}. To resolve this degeneracy only other instruments providing complementary measurements may help. In the next two sections, we will explore how this can be achieved by the VIRTIS and MIRO instruments.\\

Because Rosetta spent $\sim 65\%$ of the time at phase angles between $80-105^\circ$ and an additional $\sim 25\%$ of the time at phase angles between $50-70^\circ$ it could be argued that our analysis above is biased to phase angles not covered by Rosetta. We therefore present the corresponding plots of the number density comparison for these phase angle ranges in the appendix (Figs.~\ref{fig:rosina-lowQ_LP_50-70}-\ref{fig:rosina-highQ_SP_80-100}). These results show that for the mentioned phase angle ranges the results above are strengthened because the differences between the models are even smaller.

\subsection{The VIRTIS-M-IR view}
\begin{figure}[ht]
  \includegraphics[width=\linewidth]{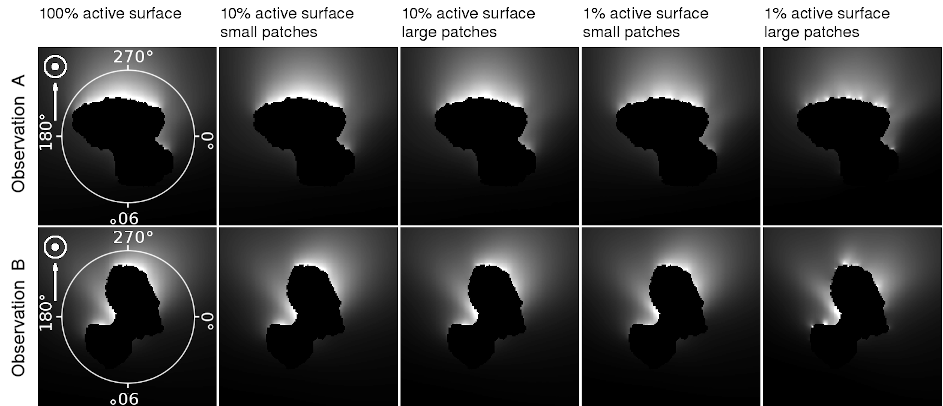}
  \caption{Synthetic VIRTIS-M-IR cubes for the five different models with a low global gas production rate for two different viewing geometries (A and B) at a phase angle of $60^\circ$.}
  \label{fig:virtis-m-ir_panel_lowQ} 
\end{figure}

A first instrument that could potentially lift the degeneracy is the VIRTIS instrument in its M-IR mapping mode. This instrument can provide image cubes of the gas coma resulting in spatially resolved maps of the gas column density \cite[see e.g.][]{Migliorini2016}. Figures~\ref{fig:virtis-m-ir_panel_lowQ} and \ref{fig:virtis-m-ir_panel_highQ} show synthetic VIRTIS-M-IR cubes for two viewing geometries (observations A and B) for the five different models with low (Fig.~\ref{fig:virtis-m-ir_panel_lowQ}) and high (Fig.~\ref{fig:virtis-m-ir_panel_highQ}) global gas production rates. The results from both synthetic observational geometries with a phase angle of $60^\circ$ are representative of the results of other viewing geometries and can thus be used to come to general conclusions.

\begin{figure}[ht]
  \includegraphics[width=\linewidth]{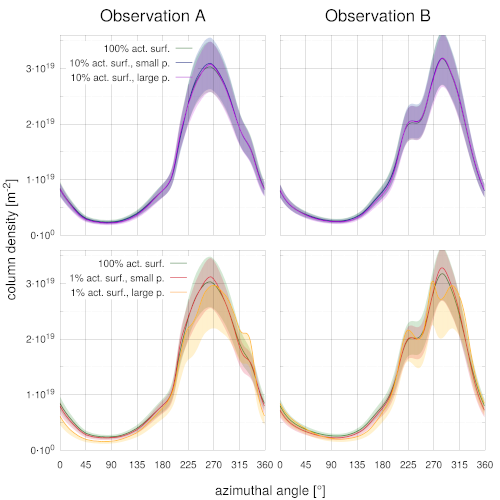}
  \caption{Gas column density of the two observing geometries A and B for the $100\%$ active area (green), $10\%$ active area (blue [small patches] and purple [large patches]), $1\%$ active area (red [small patches] and orange [large patches]) model as a function of the azimuth angle along the circles at 3~km from the nucleus centre indicated in Fig.~\ref{fig:virtis-m-ir_panel_lowQ} for a low global gas production rate. The bands indicate $\pm 15\%$ error intervals.}
  \label{fig:virtis-m-ir_azimuth_lowQ} 
\end{figure}

\begin{figure}
  \includegraphics[width=\linewidth]{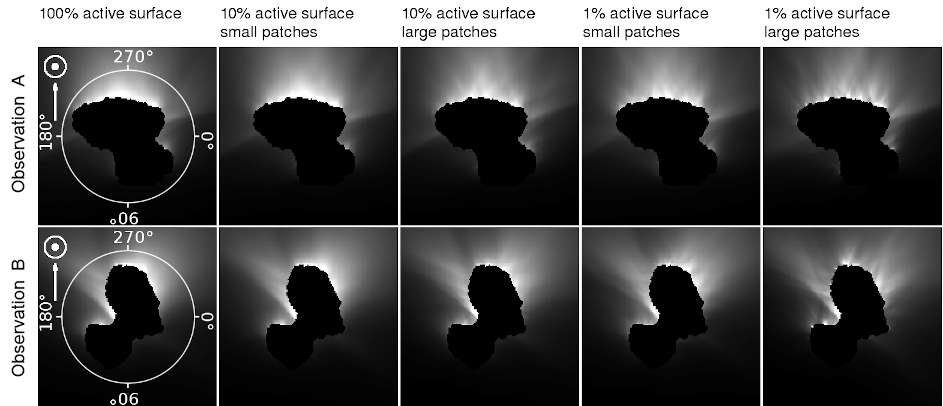}
  \caption{Synthetic VIRTIS-M-IR cubes for the five different models with a high global gas production rate for two different viewing geometries (A and B) at a phase angle of $60^\circ$.}
  \label{fig:virtis-m-ir_panel_highQ} 
\end{figure}

Looking at the low activity models presented in Fig.~\ref{fig:virtis-m-ir_panel_lowQ} several things are of importance. First, all cubes show a very smooth inner coma with little fine-scale structure. This is a reflection of the low gas flux which in turn results in large MFP and large $Kn_p$ and therefore little interaction between the molecules. As seen in Sec.~\ref{sec:theory-jets} we are still in the free penetration regime. This efficiently smoothes out any structures that could result e.g. from the surface topography or inhomogeneity of the emission due to the patches. Second, qualitatively the different models result in the same cubes and show no clear difference in the structures of the inner coma. Any inhomogeneities from the patchy emission at the surface are efficiently smoothed within the first tens to hundreds of meters above the surface. The only exception to this is the most extreme case with the activity coming from large patches from $1\%$ of the surface. This is the only case where we see a slight qualitative difference to the other models. In this case, the observations close to the limb still reveal the patchy nature of the emission. We have expected this enhancement of contrast resulting from the transition from the free to the disturbed penetration regime (see Sec.~\ref{sec:theory-jets}). Third, the two different viewing geometries show no significant difference in general behaviour. In particular, observation B with a view of the "Neck" does not show a strong focusing of the flow in the "Neck" as might be expected. The flow is somewhat collimated but with rather low contrast. For a more qualitative assessment Fig.\ref{fig:virtis-m-ir_azimuth_lowQ} shows the normalised column density as a function of the azimuth angle (AZ) along a circle at an impact parameter of $3$~km as drawn in Fig.~\ref{fig:virtis-m-ir_panel_lowQ}. Notably, the main emission direction is in both cases centred at an AZ of roughly $270^\circ$ which corresponds to the direction of the sub-solar point. The shoulder seen in observation B at an AZ of $225^\circ$ reflects the mild focusing of the flow in the "Neck". These plots also confirm the qualitative assessment above. For both observations A and B there is no detectable difference between the $100\%$ active surface (green lines) and the two $10\%$ active surface cases (blue lines) irrespective of the patch size. For the two $1\%$ active surface cases we only see deviations for the model with emission from large patches (orange lines) in particular for observation B. Therefore the right observational condition is important to potentially detect differences. We can furthermore conclude that for low global production rates only the most extreme case ($1\%$ active surface in large patches) can possibly be detected (as was true for ROSINA/COPS) and thus does not provides stronger constraints than ROSINA/COPS alone. This is especially true because of the large error bars we have assumed($\pm15\%$) in accordance with \cite{Migliorini2016}.

\begin{figure}
  \includegraphics[width=\linewidth]{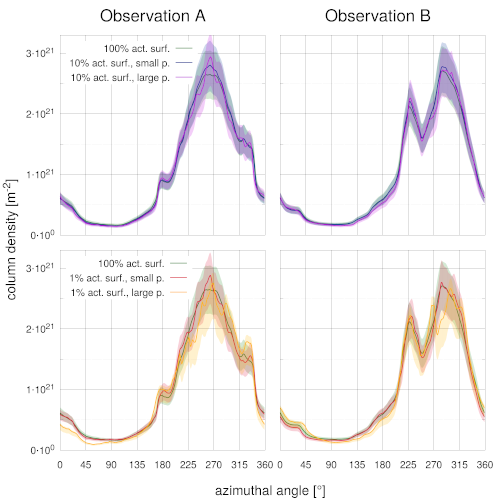}
  \caption{Gas column density for a high global gas production rate. See caption of Fig.~\ref{fig:virtis-m-ir_azimuth_lowQ}}
  \label{fig:virtis-m-ir_azimuth_highQ} 
\end{figure}

The high global gas production rate cases do not present a more optimistic picture, though as shown in Fig.~\ref{fig:virtis-m-ir_panel_highQ} we see qualitative differences. Compared to the low activity cases stronger variations in the inner coma structures can be observed in $1\%$ active surface models. Furthermore, we see clearer trends in both observations that the contrast, and thus the observable structures in the coma, increases as the fraction of area decreases and as we go from small to large patches. Whereas the $100\%$ active surface model shows almost no structure as seen in the low activity cases, the $1\%$ active surface model with large patches shows very distinct features. Not only do we see structures close to the limb but these features persist farther out. This enhanced contrast compared to the low activity cases and contrast trend is in line with what we expect (see Sec.~\ref{sec:theory-jets}). We can observe the manifestation of shocks. In these cases, we now also see strong focussing in the "Neck" with high contrast to adjacent areas. This is also reflected in the azimuthal profiles presented in Fig.~\ref{fig:virtis-m-ir_azimuth_highQ}. In particular, the focusing of the flow in the "Neck" is now strongly represented with a peak at an AZ of $225^\circ$. But due to the large error bars, it is still not possible to make a clear distinction between the different models apart again from the most extreme case that shows slight deviations.

There is thus no strong improvement over the picture seen above with the low activity models. VIRTIS-M-IR can thus not provide a tool to differentiate between different models that ROSINA/COPS cannot distinguish. Furthermore, the cryocooler of VIRTIS-M-IR failed around the time of the inbound equinox in May 2015 which was several months pre-perihelion and at a time when the activity was still rather low. After that VIRTIS-M-IR no longer provided any data and hence we have no measurements for high activity cases. But we can state that these kinds of measurements during a future mission could be an important diagnostic tool to study the inner gas coma if the accuracy can be increased from $15\%$ to below $5\%$.

\subsection{The MIRO view}
The MIRO instrument, through its resolution of individual spectral lines, is sensitive to the gas properties along the line-of-sight (LOS) of the instrument's beam. Here we consider only central LOS and will not explore effects caused by the beam shape. In particular, MIRO is sensitive to the local gas number density, speed along the LOS, and the temperature. As shown in \cite{marschall2019} the MIRO spectrum can be inverted to resolve the height profiles of these gas properties and thus be used to study the distribution of sources and the dynamics from the spacecraft down to the surface. MIRO is, therefore, an intriguing candidate for us to probe the differences of the models in this work. Should MIRO be able to detect the different models at low gas activity it follows that it will be able to do the same in the high activity cases too. Unless the observed lines are completely saturated then detecting these differences would not be possible. We, therefore, discuss here only the model results from the low activity case to show the high sensitivity of MIRO.

\begin{figure}
  \includegraphics[width=\linewidth]{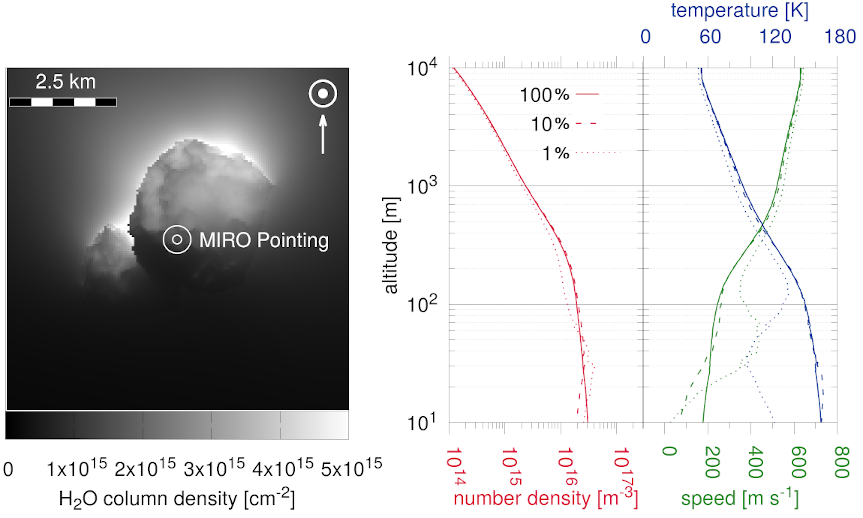}
  \caption{The left panel shows a context image of the synthetic water column density and the pointing of the MIRO beam. The two right panels show line of sight (LOS) profiles of the gas number density (red), temperature (blue), speed along the LOS (green) as a function of altitude above the comet surface for a synthetic MIRO geometry at a phase angle of $60^\circ$. The three models of varying active area ($100\%$ solid lines, $10\%$ dashed lines, $1\%$ dotted lines) for a low global gas production rate with small patches are compared. }
  \label{fig:miro_lowQ_SP} 
\end{figure}

\begin{figure}
  \includegraphics[width=\linewidth]{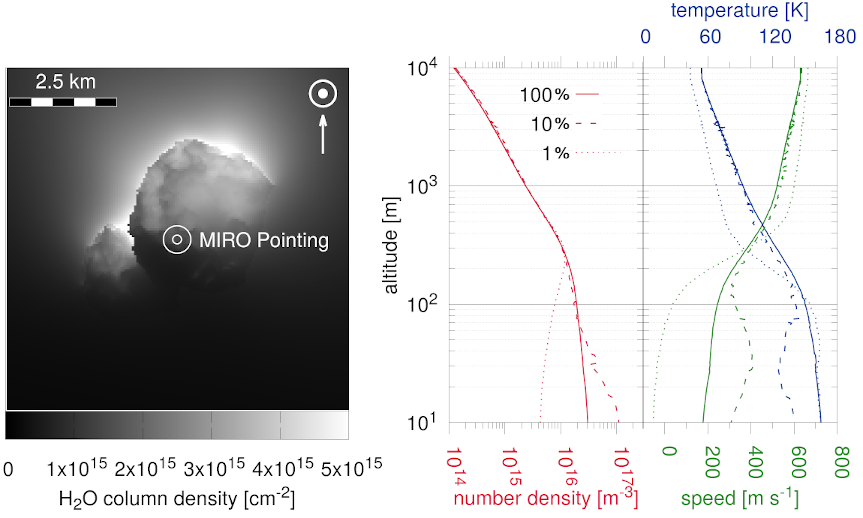}
  \caption{Shows the same properties as Fig.~\ref{fig:miro_lowQ_SP} but for large patches.}
  \label{fig:miro_lowQ_LP} 
\end{figure}

Figure~\ref{fig:miro_lowQ_SP} compares the LOS profiles of the different surface fractions with small patches at low gas activity. The profiles of the number density reveal that all models show almost identical curves which is consistent with the inability of VIRTIS and ROSINA/COPS in detecting these differences. But this does not hold for the profiles of the speed along the LOS and the temperature. While the profiles of the $100\%$ and $10\%$ active areas are very similar we see strong deviations of both of them in the case of the $1\%$ active area. The deviations, for example in the speed, will be reflected in the Doppler shift of the spectral lines seen by MIRO. Where VIRTIS and ROSINA/COPS are blind MIRO can detect clear differences in a first case.

Turning from small to large patches (Fig.~\ref{fig:miro_lowQ_LP}) makes an even stronger case. While the density profiles converge at greater than $200$~m above the surface they are distinctly different in all three active area fractions at lower altitudes. These differences are potentially strong enough to be detected in the depth of the MIRO water absorption lines H$_2^{16}$O and H$_2^{18}$O. Furthermore the profiles in temperature and speed along the LOS are also distinctly different for all three models - in the case of the $1\%$ active surface even up to high altitudes. It is also notable (seen in both figures) that we can observe temperature decreases followed by increases and again by decreases. This is likely caused by crossing and/or converging flows from different origins at the surface. From the five models shown in these two figures representing the models with low activity, notably, only two of them cannot be disentangled (the $100\%$ case and $10\%$ case with small patches). The examples shown here are representative of the majority of profiles analysed for other phase angles and viewing geometries. We should mention though that there are viewing geometries that are unsuited for this analysis where it is not possible to distinguish the models. 

For completeness and transparency the corresponding results of the high activity cases are shown in the figures in the appendix for the small (Fig.~\ref{fig:miro_highQ_SP}) and large (Fig.~\ref{fig:miro_highQ_LP}) patches corroborate the observations made above. Finally, while ROSINA/COPS and VIRTIS provide very important constraints on the location of the volatile mass loss and a good estimate of the global gas production rate they can do so only to a quite limited spatial resolution. These results illustrate that MIRO is probably the only Rosetta instrument able to resolve the degeneracy in the gas source distribution at the surface from low to high production rates. The retrieval of the discussed features by MIRO remains challenging though -- as discussed further in e.g. \cite{Marschall2019,Rezac2019} -- because the MIRO beam has a non-negligible spatial extent.

\subsection{The effect of regional inhomogeneity} \label{sec:regional-inhom}
\begin{figure}
  \includegraphics[width=\linewidth]{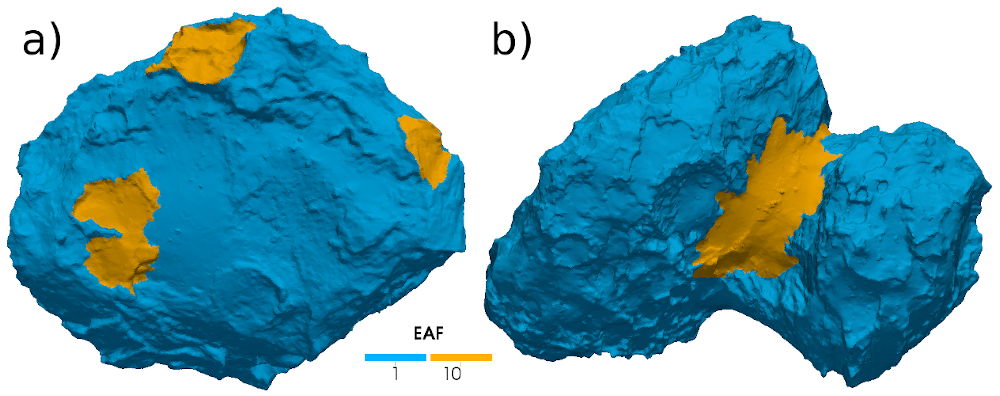}
  \caption{The two regional inhomogeneous maps are shown. a) corresponds to the inhomogeneous Imhotep map and b) to the inhomogeneous Hapi map. The highly active regions are a factor of 10 more active than the rest of the nucleus.}
  \label{fig:eaf-inhom} 
\end{figure}

Up to this point, we have only examined the effect of granularity of uniform ice distributions. But large scale regional inhomogeneity has been found on comet 67P. In this section, we will show that the drawn conclusions from above equally hold in the case of regional inhomogeneity.\\
To examine this we have overlain the maps described above and shown in Fig.~\ref{fig:eaf-all} with regional inhomogeneity. For this, we have chosen two inhomogeneity maps shown in Fig.~\ref{fig:eaf-inhom}. First, the inhomogeneous Imhotep map (shown in panel a) of Fig.~\ref{fig:eaf-inhom}) is characterised by three regions (in orange) in and around Imhotep that are a factor of 10 more active than the entire rest of the nucleus (in blue). Second, the inhomogeneous Hapi map has been inspired by our previous work \citep{Marschall2016,Marschall2017} where we have found that Hapi (orange) has a higher EAF than the rest of the comet (blue) by a factor of 10. The sub-solar longitude for this second map was chosen to be $260^{\circ}$. Both of these maps have been convoluted with the one in Fig.~\ref{fig:eaf-inhom} to create inhomogeneous patched maps. As before for high and low global gas production rates five models have been run: A $100\%$ active surface model with the regional inhomogeneity of the Imhotep or Hapi map; $10\%$ active surface models with large and small patches; and $1\%$ active surface models with large and small patches. Therefore a $1\%$ active surface model with large patches now consists of $99\%$ inactive area, and large patches in the inhomogeneous regions that are ten times more active than their counterparts everywhere else.

\begin{figure}
  \includegraphics[width=\linewidth,trim={0 0 0 2.3cm},clip]{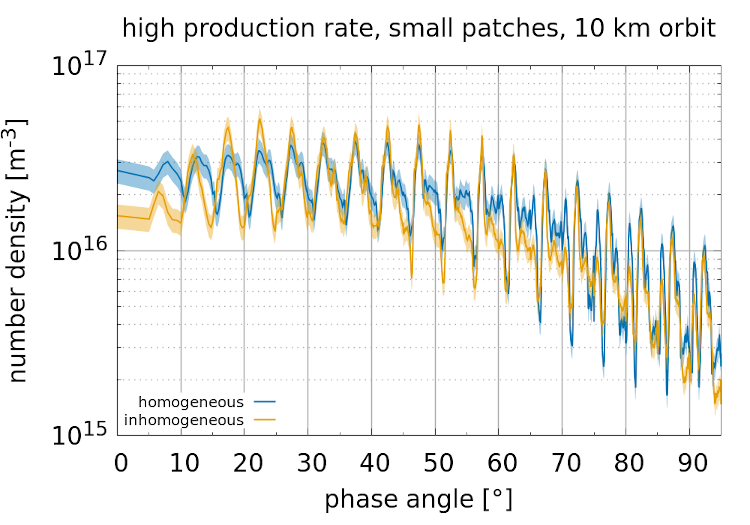}
  \caption{Local gas number density [m$^{-3}$] as a function of phase angle comparing the high production rate cases of the 100\% active surface models with homogeneous (blue) and Imhotep inhomogeneous (orange) ice distribution. The bands indicate $\pm 15\%$ error intervals.}
  \label{fig:rosina-hom-vs-inhom-highQ} 
\end{figure}

First, Fig.~\ref{fig:rosina-hom-vs-inhom-highQ} shows the comparison of the gas number density in the high activity case where $100\%$ of the surface is active one with the uniform (blue) and Imhotep inhomogeneous (orange) ice distribution. The equivalent for low activity is shown in Fig.~\ref{fig:rosina-hom-vs-inhom-lowQ}. Clear differences can already be seen qualitatively. This is backed up by the fact the mean relative difference is $31.3\%$ ($21.5\%$ for the low activity case) and the PPMCC is $0.91\pm0.01$ ( $0.95\pm0.01$ for the low activity case). We have seen above that mean relative difference $>15\%$ and PPMCC $<0.97$ can be picked up by ROSINA/COPS. Therefore this kind of regional inhomogeneity can be picked up by Rosetta measurements. This in of itself is not new. \cite{Marschall2016} found this to be true for the analysis of data in the early phase of the mission. But the question remains can differently patched inhomogeneous models be differentiated from each other.

\begin{figure}
  \includegraphics[width=\linewidth]{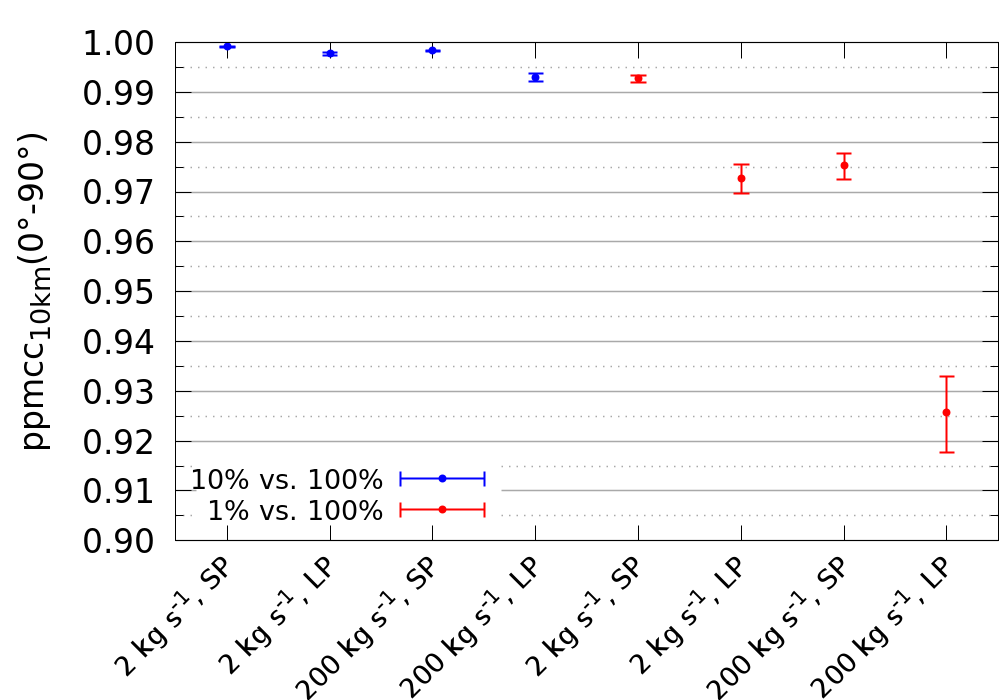}
  \caption{Pearson product-moment correlation coefficient (PPMCC) for the eight patchy Imhotep inhomogeneous cases comparing the 10\% (blue), and 1\% (red) active surface cases with small patches (SP) and large patches (LP) to the 100\% active surface case. The error bars represent a 2$\sigma$ confidence interval. The values have been calculated with the number densities at $10$~km and for phase angles covering $0-90^\circ$.}
  \label{fig:rosina-ppmcc-Imhotep} 
\end{figure}

\begin{figure}
  \includegraphics[width=\linewidth]{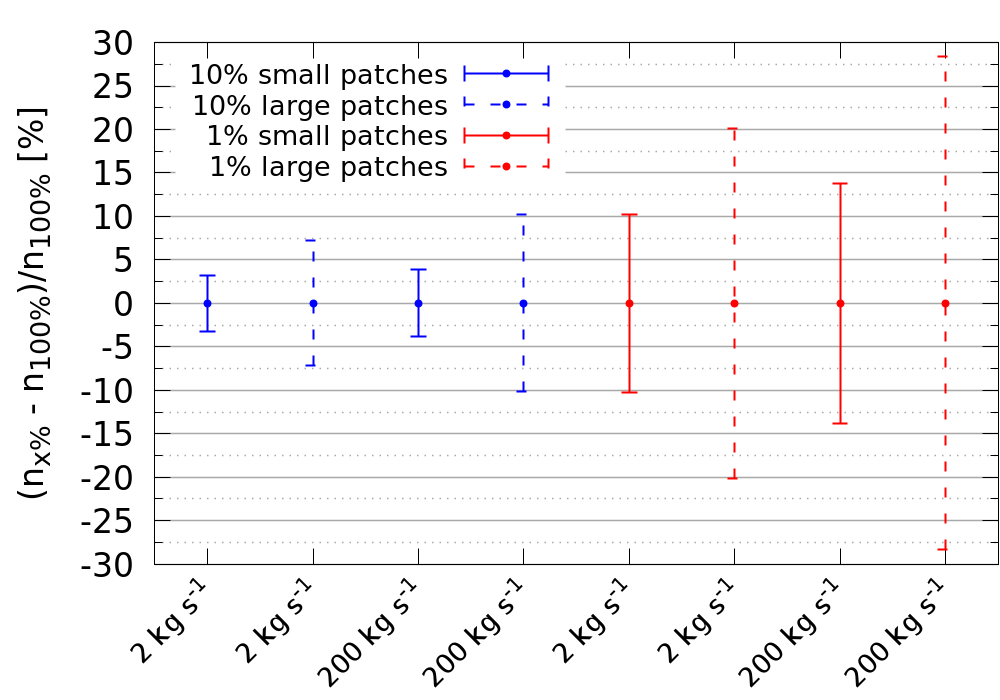}
  \caption{Mean relative difference between the number density of the eight patchy Imhotep inhomogeneous cases comparing the 10\% (blue), and 1\% (red) active surface cases with small patches (solid lines) and large patches (dashed lines) to the 100\% active surface case. The error bars represents the standard deviation of the number density differences.}
  \label{fig:rosina-delta-Imhotep} 
\end{figure}

Figs.~\ref{fig:rosina-ppmcc-Imhotep} and \ref{fig:rosina-delta-Imhotep} show the corresponding plots of the PPMCC and mean relative difference as presented above in Figs.~\ref{fig:rosina-ppmcc} and \ref{fig:rosina-delta} for the uniform ice distribution. A comparison of these plots shows that the fact that there is now an underlying inhomogeneous ice distribution does not significantly alter the correlation and mean difference between the different patchy models. The trends are the same. Except for the most extreme case of $1\%$ active surface with activity only from large patches all other cases cannot be distinguished from each other. It thus does not matter whether the underlying ice distribution is uniform or regionally inhomogeneous, ROSINA/COPS is not able to differentiate between the models of different granularity. The resolution limit remains. This holds as well for the Hapi inhomogeneous map. The corresponding results of PPMCC and mean relative difference are presented in Figs.~\ref{fig:rosina-ppmcc-Hapi} and \ref{fig:rosina-delta-Hapi}. The patchy Hapi models are closer to the non-patchy model than in the Imhotep case. This is likely caused by the fact that the emission originates in the "Neck" and thus is more confined and has less physical space to expand. Even the most extreme case is only on the cusp of detectability. This illustrates the fact that the complex shape can further complicate the determination of sources. For completeness the number density comparisons between the different models are presented in Figs.~\ref{fig:rosina-highQ_SP-Imhotep}-\ref{fig:rosina-lowQ_LP-Imhotep}. That our results hold for inhomogeneous distributions is again not surprising. A similar case has been examined by \cite{Marschall2017}. They found that an inhomogeneous distribution including emission from Hapi and cliffs could not be distinguished from a model with activity from Hapi and low level of activity everywhere else. Because cliffs are almost uniformly distributed over the surface this results in a kind of patchiness similar to that studied here.

\subsection{Simplified models}
Models assuming collision-less/free molecular flow have been among the first to be used for describing cometary comae, with the Haser model being the most prominent \citep{Haser1957}. Such simplified models are attractive in their application in comparison to DSMC models because the latter is very computationally expensive. The temptation to simplify the gas dynamics/kinetics problem is thus tempting. On the other hand, any simplified model needs to ensure that it is a correct parametrization of the necessary physics, i.e. it needs to demonstrate that it yields physically adequate and precise results. The Haser model which assumes radial outflow at a constant gas speed is clearly inadequate to describe the structure of the inner coma of a comet such as 67P. We will thus not go into detail discussing this model. Here will be looking at two other simplified models in a bit more detail. \\
\begin{figure}[h]
  \includegraphics[width=\linewidth,trim={0 0 0 2.3cm},clip]{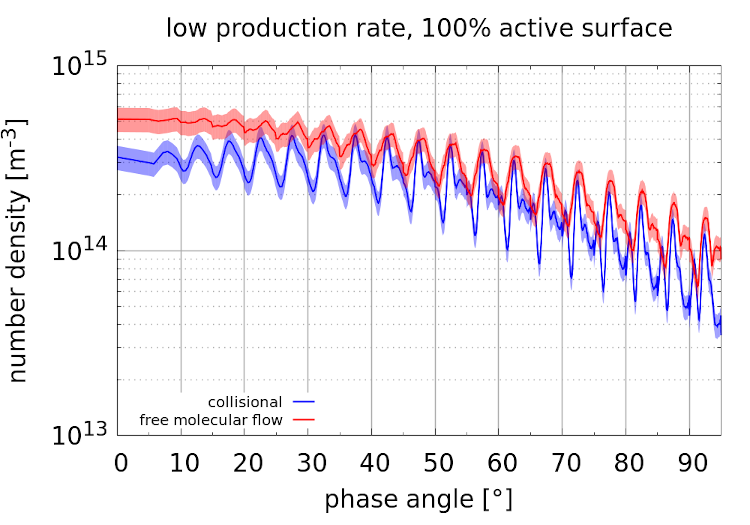}
  \caption{Local gas number density [m$^{-3}$] as a function of phase angle comparing the 100\% active surface models assuming the full physics of collisions (blue) and free molecular flow (red) for the low gas production rate.}
  \label{fig:rosina-lowQ_free} 
\end{figure}
The first model we will examine is the assumption of collisionless outflow. The larger issue with any free molecular flow model (FMFM) is that they inadequately describe the physical processes of the expanding gas. In particular, they only account for adiabatic acceleration and not for any energy and momentum exchange during collisions between molecules. To illustrate this we have run additional models with a $100\%$ active surface, and high and low gas production rates assuming no molecular collisions. The results for the predicted measurements of ROSINA/COPS for the low gas production rate case are shown in Fig.~\ref{fig:rosina-lowQ_free}. We can see that the two curves rarely overlap within the respected errors and don't show the same variation patterns. Due to the collisions the gas accelerates more in that case compared to the FMFM. In the low activity case, the maximum gas speed within our domain is 665~m/s and 465~m/s for the DSMC and FMFM respectively. For the high activity case, the speeds are 766~m/s and 603~m/s for DSMC and FMFM respectively. \\
The PPMCC of the FMFM with respect to DSMC is only $0.78 \pm 0.02$ and thus significantly below the correlations we have seen in  Fig.~\ref{fig:rosina-ppmcc} and the mean relative difference is $49.7\%$ and thus much higher than any of the models presented in Sec.~\ref{sec:ROSINA-results}. Higher production rates only make matters worse as can be seen in Fig.~\ref{fig:rosina-highQ_free}. In this case, the PPMCC is even lower at $0.68 \pm 0.03$ and the mean relative difference is even higher at $93.1\%$. It is not unsurprising that the free molecular flow approximation is not reproducing the results well, especially in the high activity case where collisions are expected to be significant. But the low activity case shows the importance of collisions even at these low production rates. We can, therefore, conclude that collisions need to be accounted for during the entirety of the Rosetta mission. Finally, though the variations of the predicted measurements are inadequately described by a collision-less model the overall order of magnitude of the expected values is of the same order of magnitude. This implies that although the spatial distribution might not be determined well, the global gas production rate estimated should not be too strongly affected by the physical simplification.

The second simplified model we examine here is the one proposed in \cite{Kramer2017}. In this model, the local gas number density at the spacecraft is the linear superposition of each facet's gas production rate scaled primarily by the inverse square of the distance from the facet to the spacecraft and the cosine of the angle between the facet normal and the direction of the spacecraft. This linear superposition facet model (LSFM), like the FMFM, does not account for collisions between molecules. This model has been used to invert local gas densities to surface production rates. When the LSFM is applied to a shape model with a large number of facets it is not clear a priori how well this model can reproduce observations given a known surface-emission map. We have therefore applied the LSFM to the all initial gas production rate maps in this work and compared the results to the physical results of our DSMC model. We have not found good agreement between LSFM and DSMC. Thus we have to conclude that the LSFM is not adequate to model the gas distribution in the inner coma and connect the surface activity to in-situ spacecraft measurements. The detailed comparison of the two models can be found in Appendix~\ref{sec:LSFM}. As with the FMFM it is plausible that the global gas production rate can be estimated to the correct order of magnitude as has been done by \cite{Kramer2017} and \cite{Laeuter2019}. On the other hand due to the analysis here it remains to be proven that LSFM can correctly connect ROSINA/COPS measurements to local gas production rates at the surface as is claimed in the above-mentioned work. Until that point the physical nature of maps presented in work using LSFM remains doubtful. Finally, to reiterate, as described in Sec.~\ref{sec:ROSINA-results} any model is physically degenerate below a spatial accuracy of a few hundred metres. This physical limitation applies irrespective of the numerical model invoked, be it an inverse or forward model. Thus no emission map can be claimed to be a physically unique solution. This also fully justifies the approach of forward modelling taken by e.g. \cite{Bieler2015,Marschall2016,Fougere2016,Zakharov2018}.

\begin{figure}[h]
  \includegraphics[width=\linewidth]{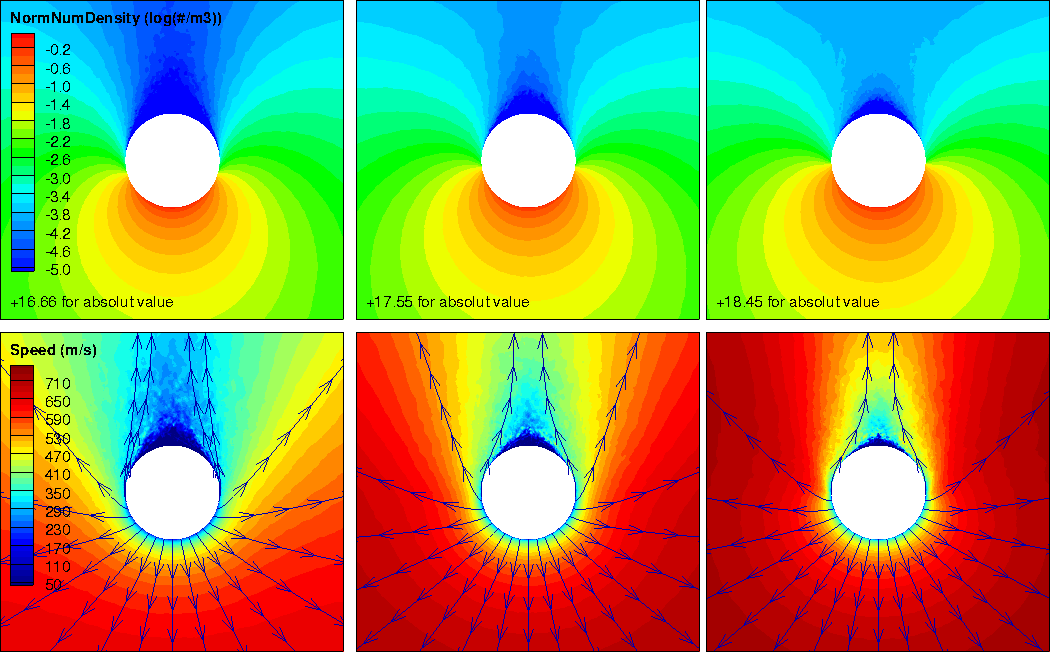}
  \caption{Slices of local gas number density (top row, [log(m$^{-3}$)]) and gas speed (bottom row, [m~s$^{-1}$]) for three global production rates of 2~kg~s$^{-1}$ (left), 20~kg~s$^{-1}$ (centre), 200~kg~s$^{-1}$ (right). A spherical nucleus with a radius of 2~km and purely insolation driven emission (i.e. globally constant EAF) is assumed. The Sun is in the bottom centre direction. The stream traces in the bottom panels originate at the same locations for each of the three production rates.}
  \label{fig:spherical-insol} 
\end{figure}

The reason why some simplifications cannot capture the structure of the inner gas coma is because the relationship between the local fluxes and in-situ measurements are in most cases non-linearly linked (due to the respective flow regime). Rather, the flow direction of a particular source depends not only on the sources' emission properties but also on the environment surrounding it. This non-locality is the main reason for the complexity in cometary gas flows.  This is illustrated even in the simplified setting shown in Fig~\ref{fig:spherical-insol}. Here we assumed a spherical nucleus with a radius of 2~km and purely insolation driven emission (i.e. globally constant EAF). Three different global gas production rates were assumed (columns from left to right: 2~kg~s$^{-1}$, 20~kg~s$^{-1}$, and 200~kg~s$^{-1}$). The sub-solar point is at the bottom centre. This figure shows that on the day side the structure in the local gas density does not vary strongly with varying gas production rate. On the night side on the other hand, as the production rate increases the densities increase relative to the day side (i.e. there is more net lateral flow from the day to the night side). On the other hand, the gas speed structure in the inner coma varies more strongly with production rate (bottom row of Fig~\ref{fig:spherical-insol}). Stream traces in the bottom panels originate at the same locations for each of the three production rates and indicate the flow path of a molecule originating at the respective surface position. These traces illustrate that molecules at low phase angles take similar paths irrespective of the production rate and can thus be linked radially with the surface. But as we go to higher phase angles the amount of lateral flow -- which depends on the production rate -- bends the stream traces resulting in non-radial paths. While in this scenario a molecule detected over the sub-solar point can reliably be connected to the sub-solar position on the surface this is not true for a molecule detected over the terminator. This is the reason why the correlation of in-situ measurements of Rosetta are particularly challenging to interpret. As mentioned above, Rosetta spend the majority of its time in orbits close to the terminator ($\sim 65\%$ of the time spent at phase angles between $80-105^\circ$).

\subsection{Lateral surface flows}\label{sec:lateral-flow}
\begin{figure}
  \includegraphics[width=0.75\linewidth]{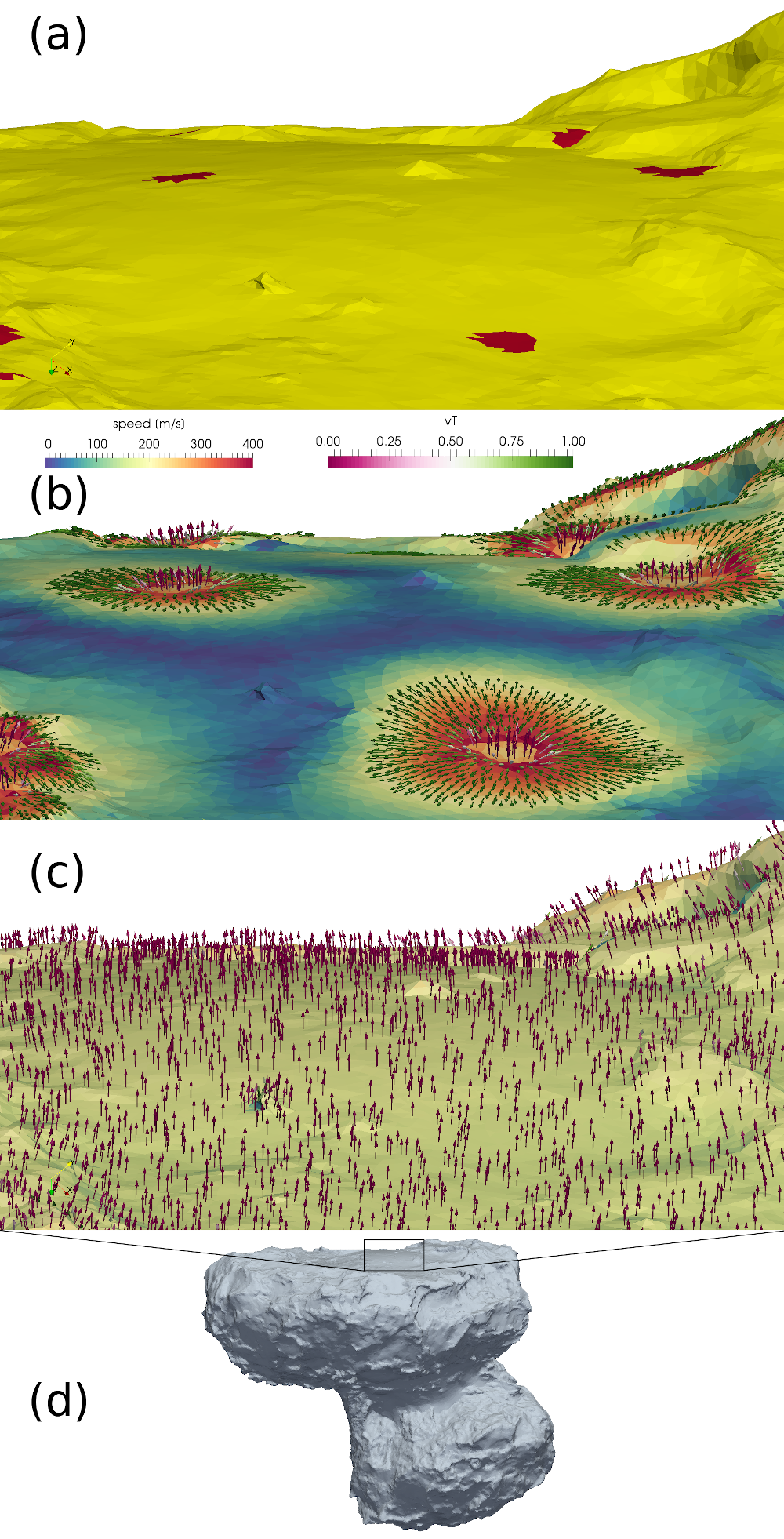}
  \caption{(a) shows the active spots in red of the large patch models, (b) for the $1\%$ active surface model with low global gas production rate the surface colour corresponds to the gas speed at the surface whereas the arrows show the direction of the flow at the surface, the arrow length is proportional to the gas speed and the colour describes the fraction of the speed that is parallel to the surface ($v_p/|\vec v|$), (c) shows the same as (b) but for the $100\%$ active surface model with low global gas production rate, and (d) gives the context of the cutout view of panels (a)-(c).}
  \label{fig:lateral-flow} 
\end{figure}

Thus far we have focused on the implications of inhomogeneous surface emission onto the inner coma and the detectability of such emission patterns by Rosetta gas instruments. In this section we want to explore some of the implications of strong variations of the outgassing on surface processes. As has been observed by \cite{Mottola2015} and \cite{LaForgia2015} the surface of comet 67P shows wind tail features and other aeolian features such as the dunes in the Hapi region \citep{Thomas15b}. In the absence of an atmosphere such features, should they be connected to sublimation processes, require rather strong lateral flows. Panels (b) and (c) of Fig.~\ref{fig:lateral-flow} show for a part of the surface the gas speed (in colour) and the arrows indicate the gas flow direction at the surface. The arrow colour additionally shows the fraction of the gas velocity that is parallel to the local surface. We can observe that a surface that has a uniform ice content as described by our $100\%$ active surface model and shown in panel (c) has a flow that is perpendicular to the surface. In such a case there are no lateral flows that could drive surface changes resulting in wind tails or dunes. In contrast a very inhomogeneous emission as described e.g. by our $1\%$ active surface large patch model shown in panel (b) results in a flow with a very different behaviour. The result is more easily understood knowing the active areas. Panel (a) shows the active areas in red. The surfaces in yellow are inert. Knowing this we can understand the flow pattern in panel (b) which illustrates that the flow over the active areas which emit gas is also perpendicular. But in the immediate vicinity the flow is on the one hand faster but already more parallel to the surface because it has less resistance as it is flowing into a lower density region than the one directly above the active areas. Only a few tens of meters away from the active region the flow is almost completely parallel to surface. This is in stark contrast to the more uniform emission seen in panel (c). We can therefore conclude that strong lateral flows can only arise in regions with a very high contrast in the strength of the emission. This was also illustrated less systematically by \citet{Thomas15b} and is somewhat intuitive.

We should state though that we do not imply from this work that the cometary activity on comet 67P originates from such large patches. The conclusion is simply that strong lateral winds can only arise in regions with abrupt change and high contrast in the emission. This behaviour is invariant over different scales and still holds e.g. if the sources are much smaller. For smaller sources, the region affected by lateral flow is simply smaller too. Features such as the wind tails, dunes and the other aeolian features need to be put in this context should one attempt to explain them as a direct consequence of sublimation and associated lateral flow.

%%%%%%%%%%%%%%%%%%%%%%%%%%%%%%%%%%%%%%%%%%%%%%%%%%%%%%%%%%%%%%%%%%%%%%
%%%%%%%%%%%%%%%%%%%%%%%%%%%%%%%%%%%%%%%%%%%%%%%%%%%%%%%%%%%%%%%%%%%%%%
%%%%%%%%%%%%%%%%%%%%    Summary and Conclusions  %%%%%%%%%%%%%%%%%%%%%
%%%%%%%%%%%%%%%%%%%%%%%%%%%%%%%%%%%%%%%%%%%%%%%%%%%%%%%%%%%%%%%%%%%%%%
%%%%%%%%%%%%%%%%%%%%%%%%%%%%%%%%%%%%%%%%%%%%%%%%%%%%%%%%%%%%%%%%%%%%%%
\section{Summary and Conclusions}
In this work, we have examined a set of "educational" models of comet 67P's gas emission. These were set up in a way to cover the most extreme cases in the distribution of sources but with equal global production rate. It was our goal to examine how different Rosetta gas instruments can distinguish between these models and have reached the following conclusions:

\begin{itemize}
    \item We have seen the influence of different flow regimes with the corresponding jet interaction regimes. In particular, the studied models covered almost the entire range from the free penetration regime to the shock interaction regime.
    \item The sensitivity of linking ROSINA/COPS measurements of the local gas density at the spacecraft to the surface distribution of emission is very low. Except for the most extreme case tested we were not able to disentangle the other models from each other.
    \item ROSINA/COPS measurements are well suited to detect larger-scale/regional inhomogeneities of the surface-emission.
    \item While ROSINA/COPS cannot determine the gas sources to a high spatial resolution it can nevertheless be used in conjunction with the appropriate physical models to estimate the global gas production rate.
    \item The degeneracy in the surface-emission distribution arising from constraints only from ROSINA/COPS is fundamental. Neither forward nor inverse models can lift this degeneracy. All model solutions fitting/constrained by ROSINA/COPS data only will be non-unique spatially at resolutions $< 400$~m ($< 50$~MFP).
    \item In general, the higher the gas production rate, the larger the patch size, and the larger the separation between patches, the easier it is to detect the sources using only ROSINA/COPS measurements.
    \item The above discussed physical limit leads to the fact that the ROSINA/COPS data cannot be inverted to give a unique surface distribution of the emission.
    \item Simplified models (such as Haser, FMFM, or LSFM), though computationally advantageous, cannot by definition adequately describe the structure of the inner gas coma.
    \item Only a multi-instrument approach with complementary measurements can lift the observed degeneracy.
    \item Measurements by VIRTIS-M-IR do not provide constraints beyond the ones provided by ROSINA/COPS.
    \item The only instrument which was able to distinguish between most cases is MIRO. Hence MIRO data provide the highest potential for improving current models of the distribution of the mass loss.
\end{itemize}

\section{Acknowledgements}

We thank Frank Preusker and Frank Scholten for providing us with the comet shape model SHAP7 \citep{Preusker2017} used in this work.\\
This work was supported through the Swiss National Science Foundation. \\
Raphael Marschall acknowledges the support of the Swiss National Science Foundation (SNSF) through the grant P2BEP2\_184482.\\
Nicolas Thomas acknowledges funding by the Swiss National Science Foundation through the NCCR PlanetS.\\
We acknowledge the personnel at ESA's European Space Operations Center (ESOC) in Darmstadt, Germany, European Space Astronomy Center (ESAC) in Spain, and at ESA for the making the Rosetta mission possible.

\appendix

\section{Linear superposition facet model (LSFM)} \label{sec:LSFM}
Here we examine how well the linear superposition facet model (LSFM) can reproduce physical results from the DSMC model. The LSFM assumes that each surface facet visible to the spacecraft can contribute to the in-situ gas number density. The contribution of each facet to the total density is calculated by scaling the facets gas production rate with the cosine of the angle between the facets surface normal and the direction of the spacecraft, $\theta$, and by the inverse of the distance from the facet to the spacecraft square, the latter accounting for the decrease of the density assuming a Haser type non-accelerated flow. The facet production rate is further scaled by $U_0 \cdot exp(-U_0^2 sin^2\theta)$. The parameter $U_0$ accounts for the degree of later expansion of the gas in the direction perpendicular to the facet normal. The density at the spacecraft is thus a linear superposition of the contributions of all facets. The full equation can be found by substituting Eq.~(1) \& (4) in Eq.~(5) of \cite{Kramer2017}. The fact that the densities at the spacecraft are simply a superposition of scaled values for each surface facet makes the LSFM very fast and thus computationally attractive.
\\
\begin{figure}[h]
  \includegraphics[width=\linewidth]{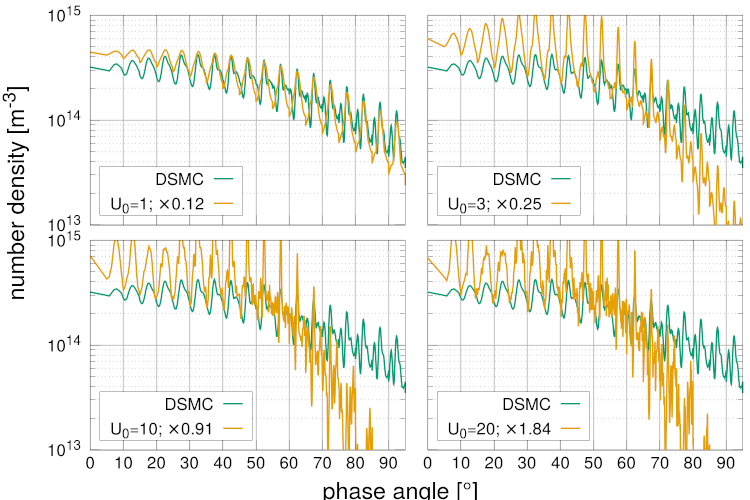}
  \caption{Local gas number density [m$^{-3}$] at 10~km from the nucleus centre as a function of phase angle comparing the low activity 100\% active surface DSMC model (green) and the LSFM (orange) for four values of $U_0$. The legend of each panel also gives the scaling factor needed to match the mean density of LSFM to the DSMC result.}
  \label{fig:KramerLaeuter-lowQ} 
\end{figure}

\begin{figure}[h]
  \includegraphics[width=\linewidth]{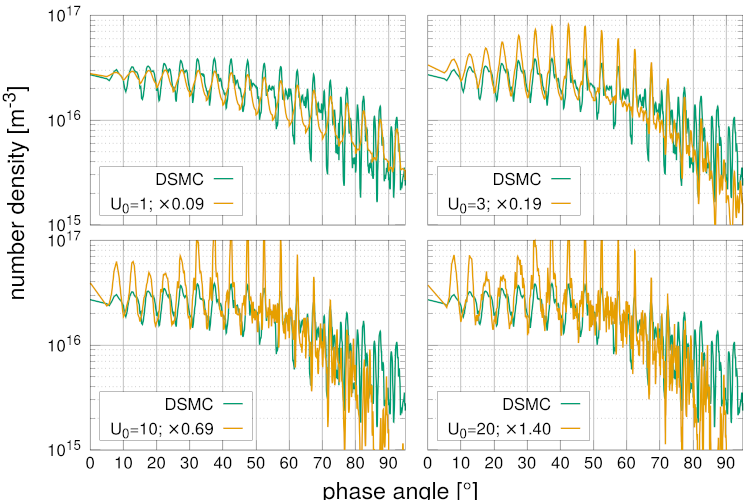}
  \caption{Local gas number density [m$^{-3}$] at 10~km from the nucleus centre as a function of phase angle comparing the low activity 100\% active surface DSMC model (green) and the LSFM (orange) for four values of $U_0$. The legend of each panel also gives the scaling factor needed to match the mean density of LSFM to the DSMC result.}
  \label{fig:KramerLaeuter-highQ} 
\end{figure}

\begin{figure}[h]
  \includegraphics[width=\linewidth]{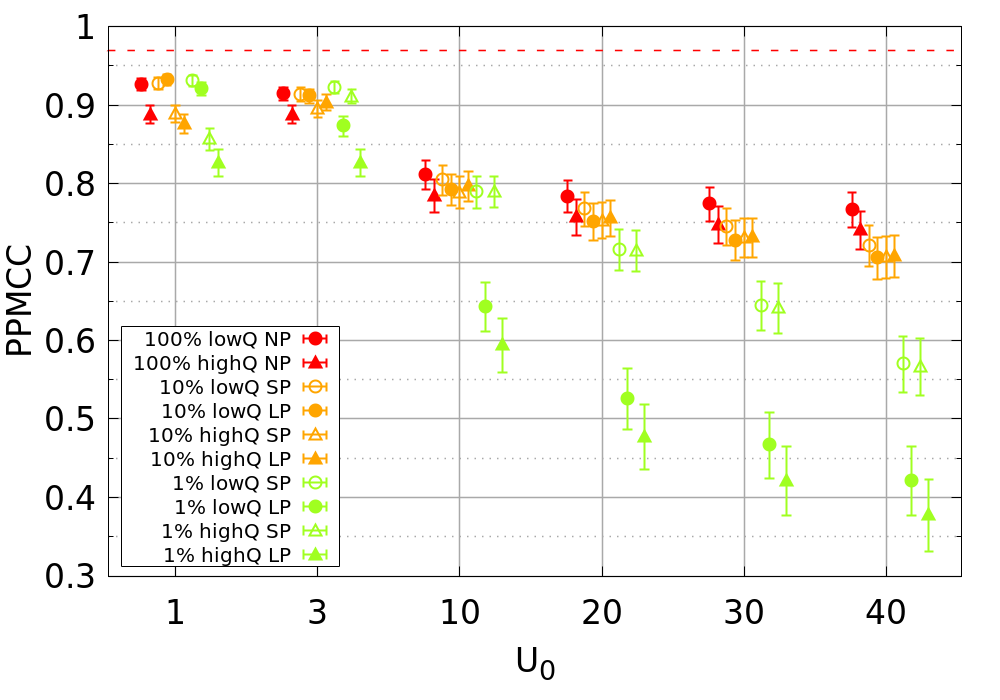}
  \caption{PPMCC of the LSFM with reference to the corresponding DSMC model for $100\%$ (red symbols), $100\%$ (orange symbols), and $100\%$ active surface (green symbols). Circles represent the low activity cases, while triangles correspond to the high activity cases. Empty symbols denote models with small patches while solid symbols denote models with large patches. }
  \label{fig:KramerLaeuter-ppmcc} 
\end{figure}

\begin{figure}[h]
  \includegraphics[width=\linewidth]{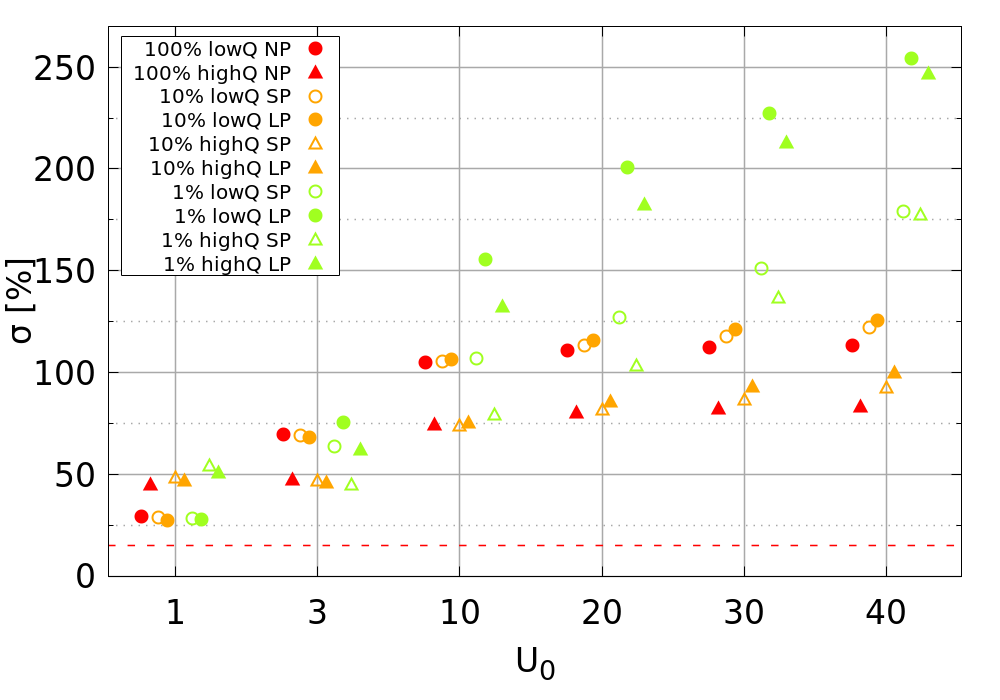}
  \caption{Mean relative deviation of the LSFM with respect to the corresponding DSMC model. See caption of Fig.~\ref{fig:KramerLaeuter-ppmcc} for symbol explanation. }
  \label{fig:KramerLaeuter-chisquared} 
\end{figure}
If the density at the spacecraft is unknown then the LSFM has $n+1$ free parameters, where $n$ is the number of surface facets: There are the gas production rates of each facet ($n$ free parameters) and $U_0$. Thus, if the surface gas production rates are known the LSFM reduces to a problem with only one free parameter ($U_0$). Here we will compare the predicted densities at the spacecraft of LSFM with the ones from our DSMC model using the same initial surface production rates. The LSFM thus still leaves one free parameter. For the choice of $U_0$ we follow the proposed values in literature. From Fig.~(4) of \cite{Kramer2017} values of $U_0 \sim 10, 30, 40$ can be calculated, while \cite{Laeuter2019} has found the best fit with a value of $U_0=3$. In addition to the above mentioned values we have also examined $U_0 = 0, 20$.\\
We have used the same ten initial conditions as described in Sec.~\ref{sec:method-and-model-setup} resulting from the maps illustrated in Fig.~\ref{fig:eaf-all}. Figures~\ref{fig:KramerLaeuter-lowQ} and \ref{fig:KramerLaeuter-highQ} shows the comparison of the resulting gas number densities at the spacecraft position for the $100\%$ active surface model resulting from the DSMC model (green lines) and the LSFM with four different values of $U_0$ (orange lines) for the low and high activity cases respectively. First, we notice that for high values of $U_0$ the variations in the number density are much larger than observed with DSMC. This is primarily due to the exponential term that suppressed contributions from high activity regions at high emission angles. This is particularly pronounced in the low activity case. In that case, only the case of $U_0=1$ reproduces the overall trend observed by in the DSMC result. In the high activity case, both $U_0=1$ and $3$ arguably reproduce the overall trend in the density. But none of the models reproduces well the small variations found in the DSMC result. This is illustrated by the quantitative assessment presented in Figs.~\ref{fig:KramerLaeuter-ppmcc} and \ref{fig:KramerLaeuter-chisquared} showing the PPMCC and relative deviation of all ten models. The LSFM results needed to be - often significantly - re-scaled (using the values in the legends) to match the average density of the DSMC results. Under the re-scaled conditions, the LSFM performs best for the low activity cases and $U_0=1$ with mean deviations to the DSMC models of $\sim 30\%$. In the worst case, the PPMCC goes as low as $0.38$ and mean deviations are as high as $250\%$. For reference, the red dashed lines denote the values for which we have found a good agreement of the patchy models with the uniform models as described in Sec.~\ref{sec:results}. We had concluded there that ROSINA/COPS is unable to detect differences between the models only, if the mean deviation is $<15\%$ and the PPMCC~$>0.97$. None of the LSFM cases satisfies these conditions and can therefore not be considered to reproduce the physical results of the DSMC model. Conversely, this also implies that we could not establish a link between the surface-emission maps and the Rosetta local gas measurements using LSFM.

\section{Supplementary figures}
The figures in this appendix complete the series of plots showing the remaining models not discussed in detail in the main text.

\begin{figure}[h]
  \includegraphics[width=\linewidth,trim={0 0 0 2.3cm},clip]{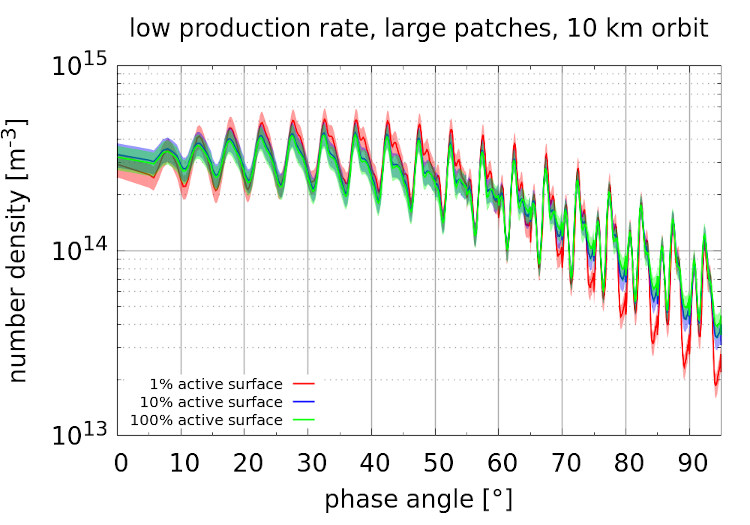}
  \caption{Local gas number density [m$^{-3}$] as a function of phase angle comparing the 100\% (green), 10\% (blue), and 1\% (red) active surface models with large patches for the low activity case.}
  \label{fig:rosina-lowQ_LP} 
\end{figure}

\begin{figure}[h]
  \includegraphics[width=\linewidth,trim={0 0 0 2.3cm},clip]{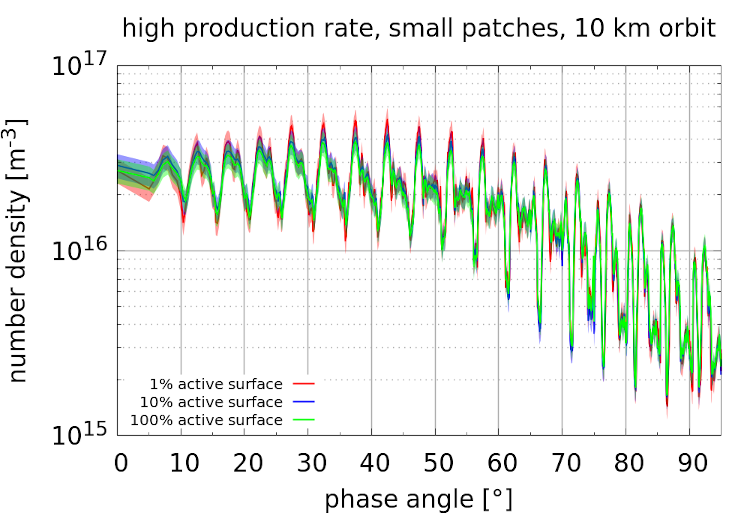}
  \caption{Local gas number density [m$^{-3}$] as a function of phase angle comparing the 100\% (green), 10\% (blue), and 1\% (red) active surface models with small patches for the high activity case.}
  \label{fig:rosina-highQ_SP} 
\end{figure}

\begin{figure}[h]
  \includegraphics[width=\linewidth,trim={0 0 0 2.3cm},clip]{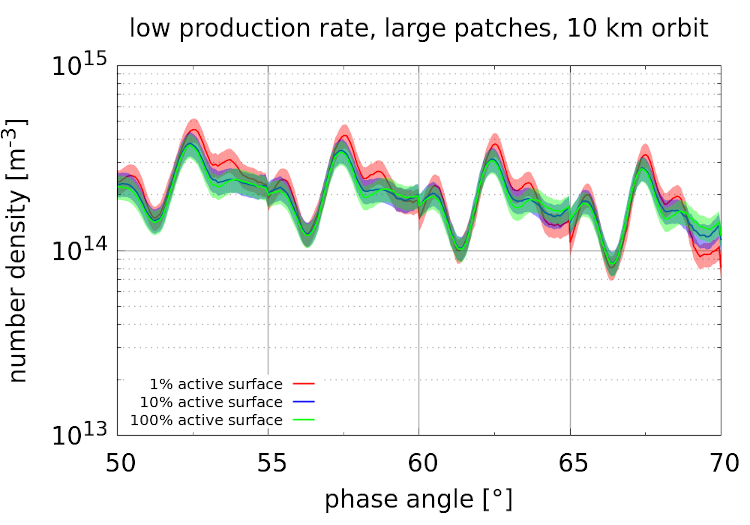}
  \caption{Local gas number density [m$^{-3}$] as a function of phase angle between $50-70^\circ$ comparing the 100\% (green), 10\% (blue), and 1\% (red) active surface models with large patches for the low activity case.}
  \label{fig:rosina-lowQ_LP_50-70} 
\end{figure}

\begin{figure}[h]
  \includegraphics[width=\linewidth,trim={0 0 0 2.3cm},clip]{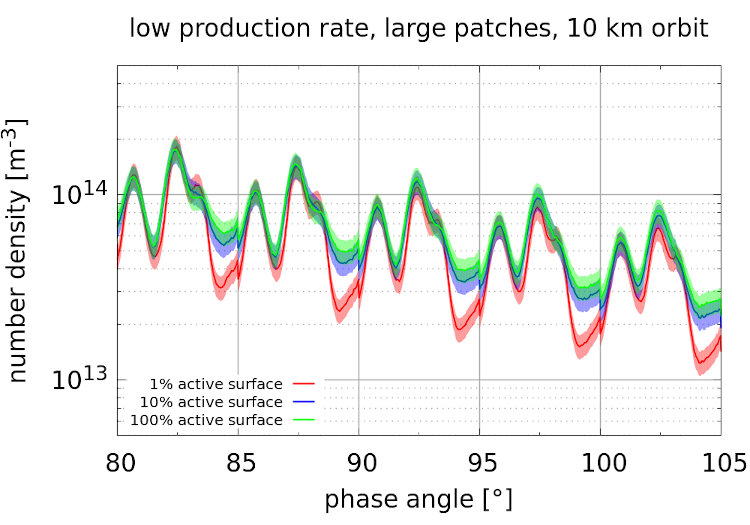}
  \caption{Local gas number density [m$^{-3}$] as a function of phase angle between $80-105^\circ$ comparing the 100\% (green), 10\% (blue), and 1\% (red) active surface models with large patches for the low activity case.}
  \label{fig:rosina-lowQ_LP_80-100} 
\end{figure}

\begin{figure}[h]
  \includegraphics[width=\linewidth,trim={0 0 0 2.3cm},clip]{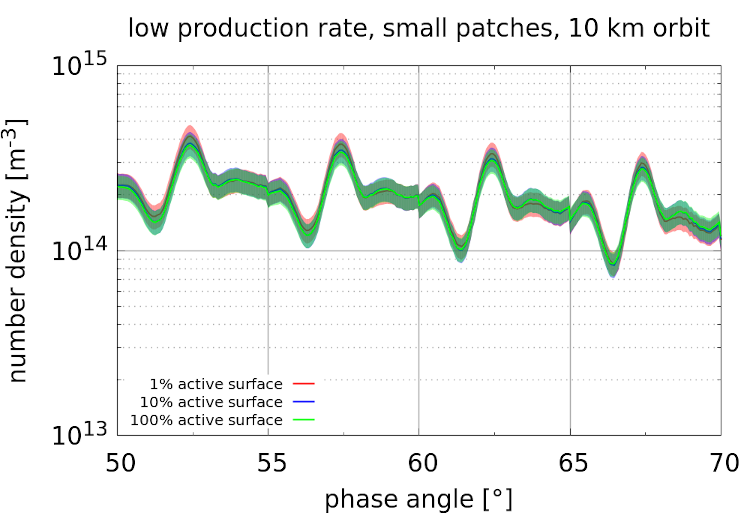}
  \caption{Local gas number density [m$^{-3}$] as a function of phase angle between $50-70^\circ$ comparing the 100\% (green), 10\% (blue), and 1\% (red) active surface models with small patches for the low activity case.}
  \label{fig:rosina-lowQ_SP_50-70} 
\end{figure}

\begin{figure}[h]
  \includegraphics[width=\linewidth,trim={0 0 0 2.3cm},clip]{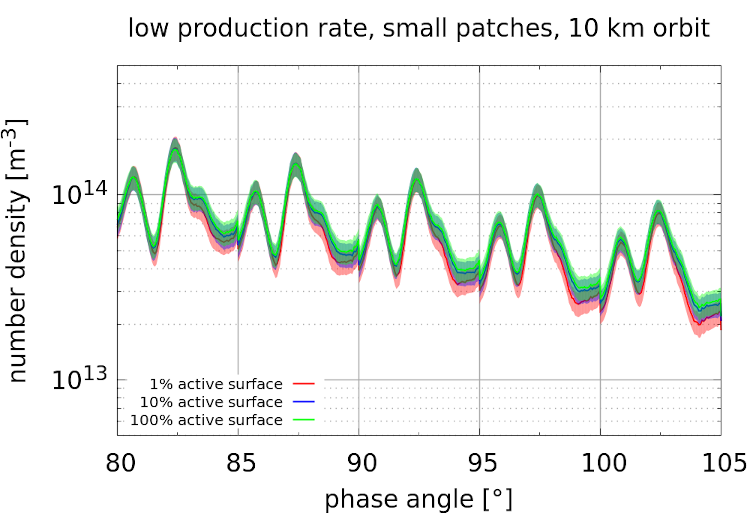}
  \caption{Local gas number density [m$^{-3}$] as a function of phase angle between $80-105^\circ$ comparing the 100\% (green), 10\% (blue), and 1\% (red) active surface models with small patches for the low activity case.}
  \label{fig:rosina-lowQ_SP_80-100} 
\end{figure}

\begin{figure}[h]
  \includegraphics[width=\linewidth,trim={0 0 0 2.3cm},clip]{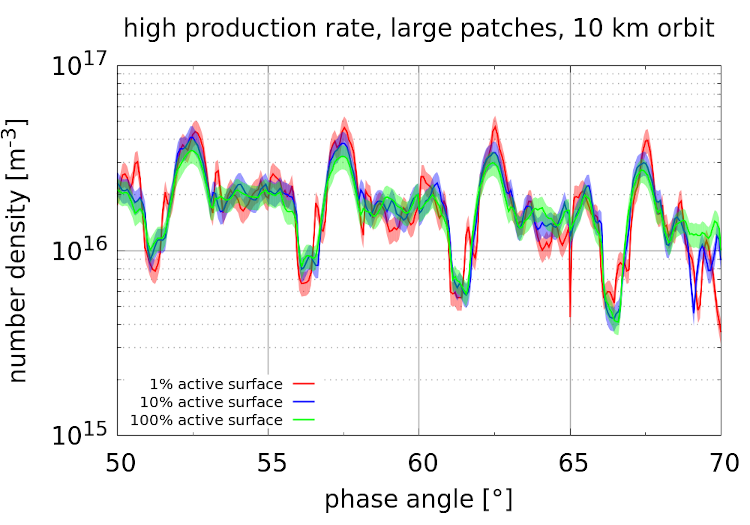}
  \caption{Local gas number density [m$^{-3}$] as a function of phase angle between $50-70^\circ$ comparing the 100\% (green), 10\% (blue), and 1\% (red) active surface models with large patches for the high activity case.}
  \label{fig:rosina-highQ_LP_50-70} 
\end{figure}

\begin{figure}[h]
  \includegraphics[width=\linewidth,trim={0 0 0 2.3cm},clip]{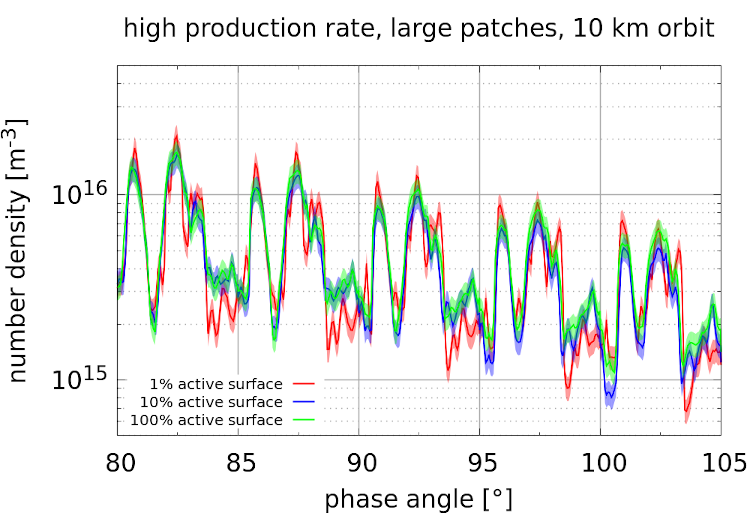}
  \caption{Local gas number density [m$^{-3}$] as a function of phase angle between $80-105^\circ$ comparing the 100\% (green), 10\% (blue), and 1\% (red) active surface models with large patches for the high activity case.}
  \label{fig:rosina-highQ_LP_80-100} 
\end{figure}

\begin{figure}[h]
  \includegraphics[width=\linewidth,trim={0 0 0 2.3cm},clip]{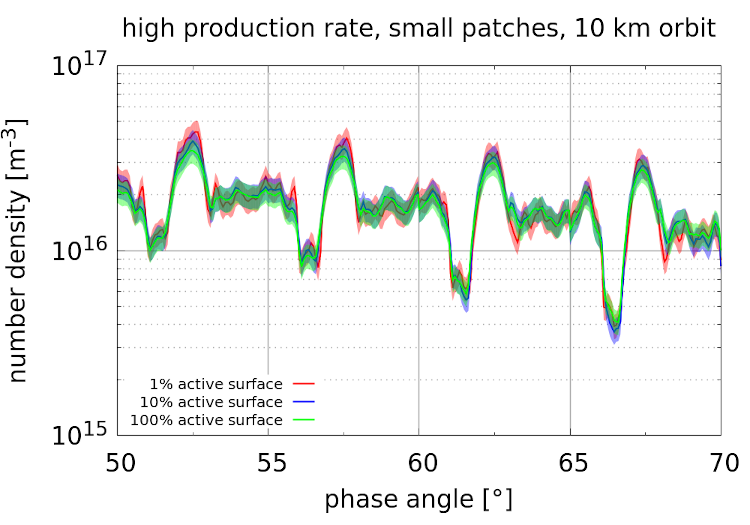}
  \caption{Local gas number density [m$^{-3}$] as a function of phase angle between $50-70^\circ$ comparing the 100\% (green), 10\% (blue), and 1\% (red) active surface models with small patches for the high activity case.}
  \label{fig:rosina-highQ_SP_50-70} 
\end{figure}

\begin{figure}[h]
  \includegraphics[width=\linewidth,trim={0 0 0 2.3cm},clip]{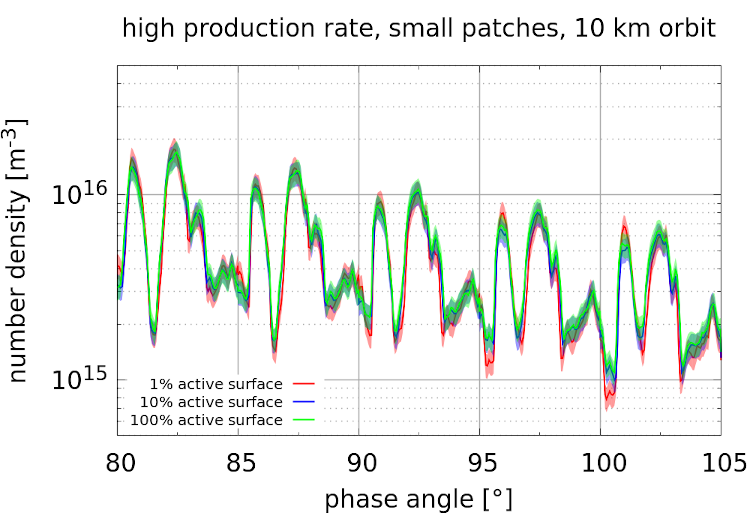}
  \caption{Local gas number density [m$^{-3}$] as a function of phase angle between $80-105^\circ$ comparing the 100\% (green), 10\% (blue), and 1\% (red) active surface models with small patches for the high activity case.}
  \label{fig:rosina-highQ_SP_80-100} 
\end{figure}

\begin{figure}[h]
  \includegraphics[width=\linewidth]{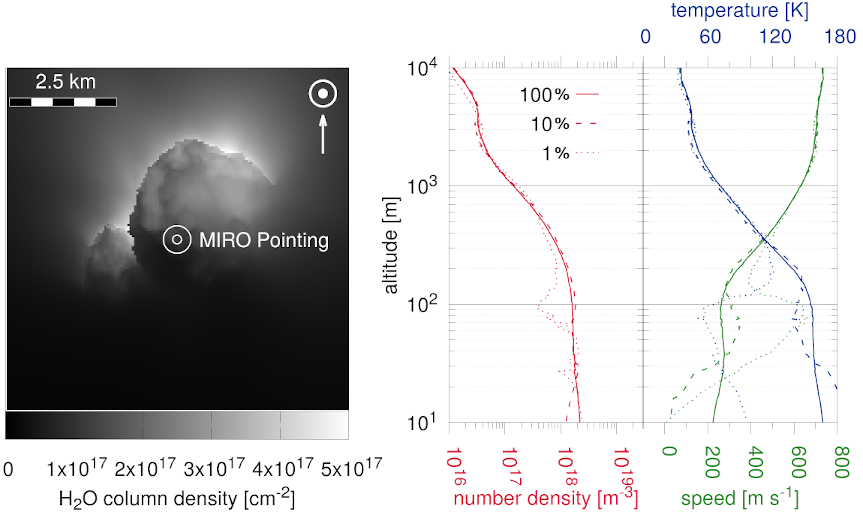}
  \caption{Shows the same properties as Fig.~\ref{fig:miro_lowQ_SP} but for small patches with high global gas production rate.}
  \label{fig:miro_highQ_SP} 
\end{figure}

\begin{figure}[h]
  \includegraphics[width=\linewidth]{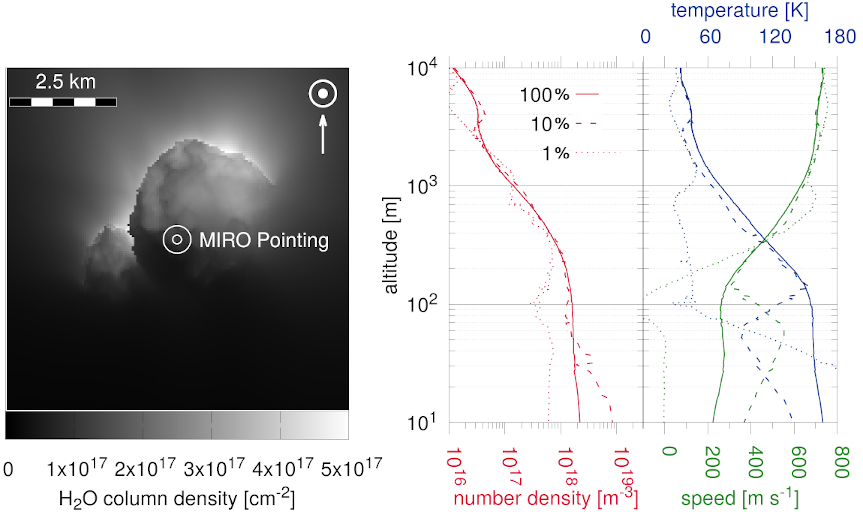}
  \caption{Shows the same properties as Fig.~\ref{fig:miro_lowQ_SP} but for large patches with high global gas production rate.}
  \label{fig:miro_highQ_LP} 
\end{figure}

\begin{figure}
  \includegraphics[width=\linewidth,trim={0 0 0 2.3cm},clip]{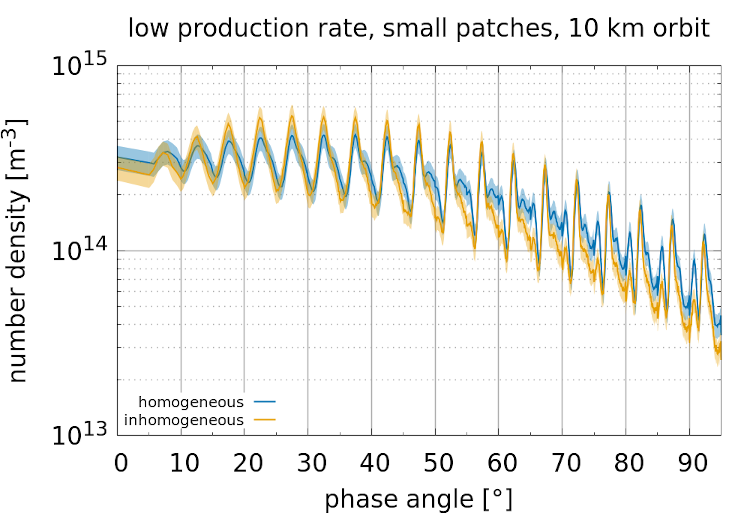}
  \caption{Local gas number density [m$^{-3}$] as a function of phase angle comparing the low production rate cases of the 100\% active surface models with homogeneous (blue) and Imhotep inhomogeneous (orange) ice distribution. The bands indicate $\pm 15\%$ error intervals.}
  \label{fig:rosina-hom-vs-inhom-lowQ} 
\end{figure}

\begin{figure}
  \includegraphics[width=\linewidth]{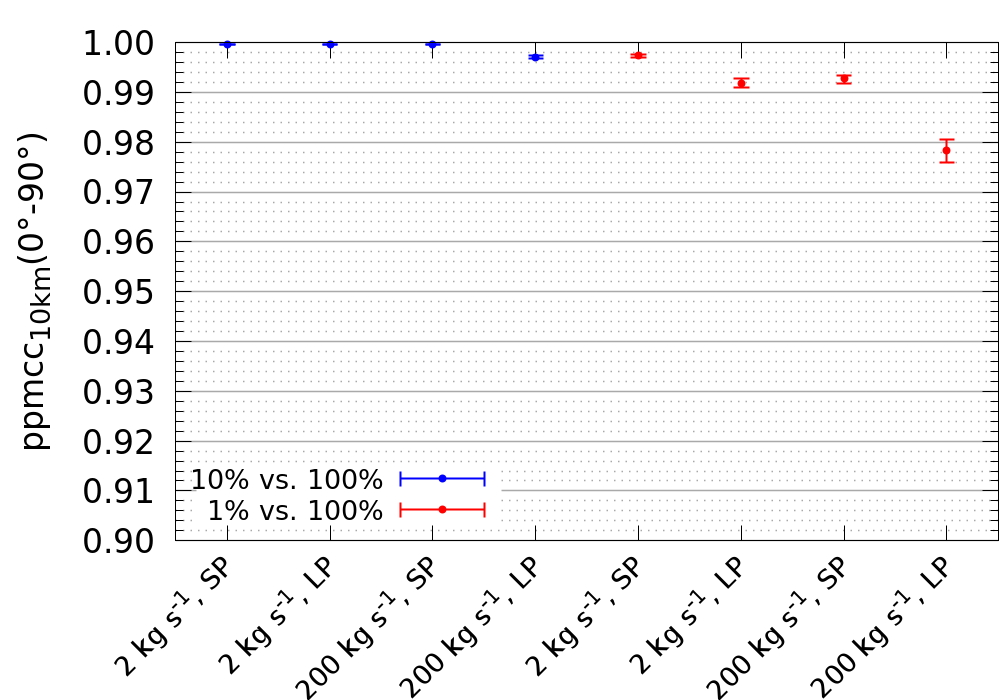}
  \caption{Pearson product-moment correlation coefficient (PPMCC) for the eight patchy Hapi inhomogeneous cases comparing the 10\% (blue), and 1\% (red) active surface cases with small patches (SP) and large patches (LP) to the 100\% active surface case. The error bars represent a 2$\sigma$ confidence interval. The values have been calculated with the number densities at $10$~km and for phase angles covering $0-90^\circ$.}
  \label{fig:rosina-ppmcc-Hapi} 
\end{figure}

\begin{figure}
  \includegraphics[width=\linewidth]{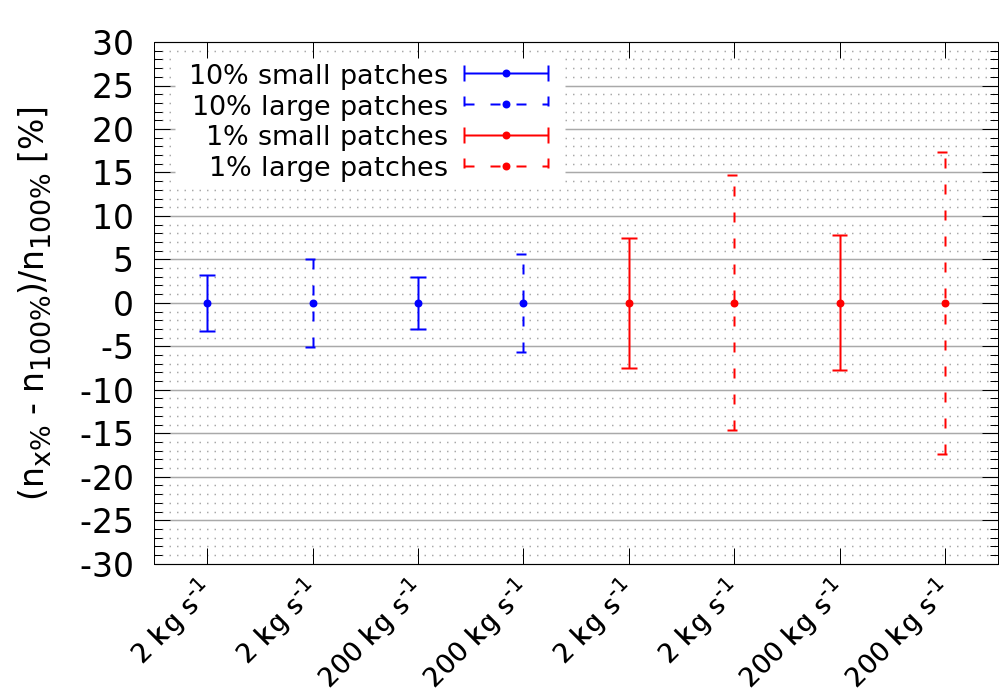}
  \caption{Mean relative difference between the number density of the eight patchy Hapi inhomogeneous cases comparing the 10\% (blue), and 1\% (red) active surface cases with small patches (solid lines) and large patches (dashed lines) to the 100\% active surface case. The error bars represents the standard deviation of the number density differences.}
  \label{fig:rosina-delta-Hapi} 
\end{figure}

\begin{figure}
  \includegraphics[width=\linewidth,trim={0 0 0 2.3cm},clip]{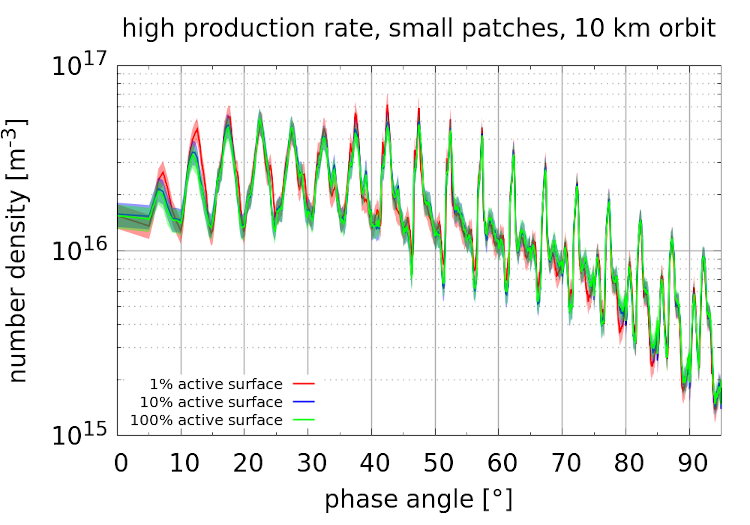}
  \caption{Local gas number density [m$^{-3}$] as a function of phase angle comparing the 100\% (green), 10\% (blue), and 1\% (red)  active surface models with small patches for the high activity case. The bands indicate $\pm 15\%$ error intervals.}
  \label{fig:rosina-highQ_SP-Imhotep} 
\end{figure}

\begin{figure}
  \includegraphics[width=\linewidth,trim={0 0 0 2.3cm},clip]{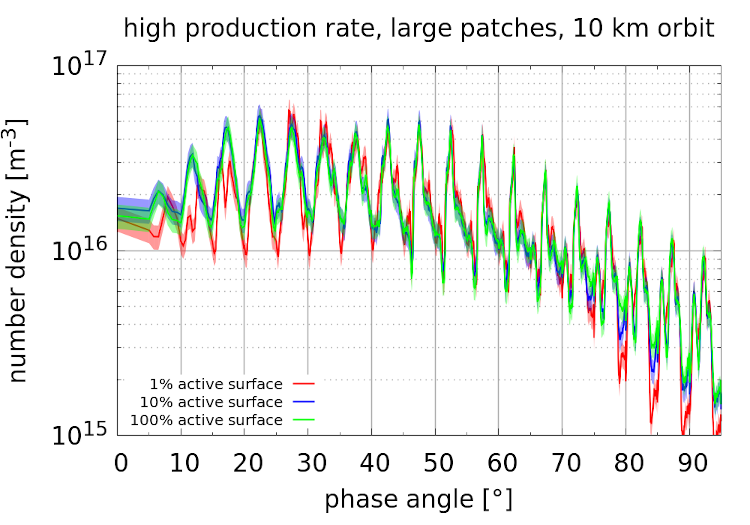}
  \caption{Local gas number density [m$^{-3}$] as a function of phase angle comparing the 100\% (green), 10\% (blue), and 1\% (red)  active surface models with large patches for the high activity case. The bands indicate $\pm 15\%$ error intervals.}
  \label{fig:rosina-highQ_LP-Imhotep} 
\end{figure}

\begin{figure}
  \includegraphics[width=\linewidth,trim={0 0 0 2.3cm},clip]{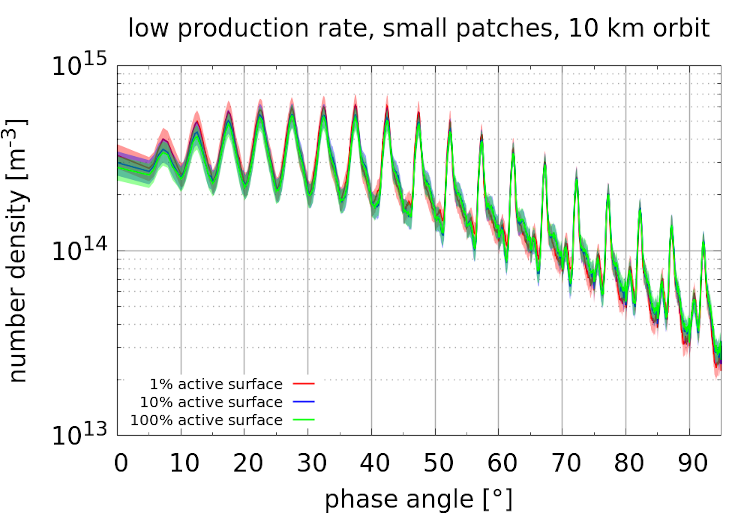}
  \caption{Local gas number density [m$^{-3}$] as a function of phase angle comparing the 100\% (green), 10\% (blue), and 1\% (red)  active surface models with small patches for the low activity case. The bands indicate $\pm 15\%$ error intervals.}
  \label{fig:rosina-lowQ_SP-Imhotep} 
\end{figure}

\begin{figure}
  \includegraphics[width=\linewidth,trim={0 0 0 2.3cm},clip]{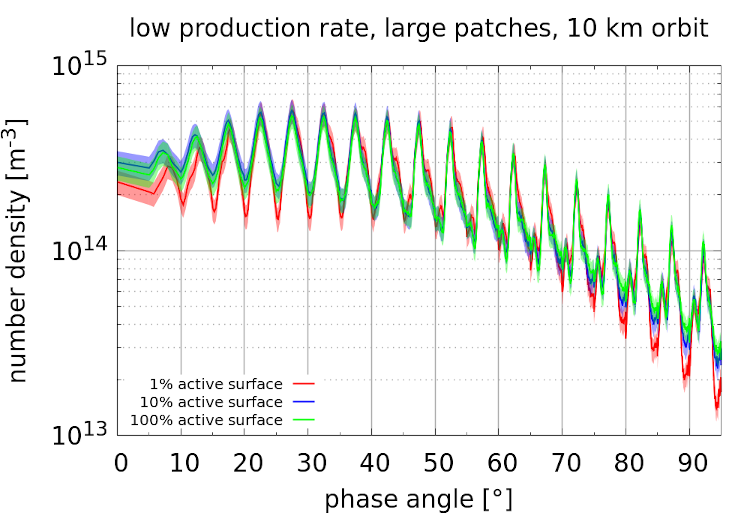}
  \caption{Local gas number density [m$^{-3}$] as a function of phase angle comparing the 100\% (green), 10\% (blue), and 1\% (red)  active surface models with large patches for the low activity case. The bands indicate $\pm 15\%$ error intervals.}
  \label{fig:rosina-lowQ_LP-Imhotep} 
\end{figure}

\begin{figure}[h]
  \includegraphics[width=\linewidth,trim={0 0 0 2.3cm},clip]{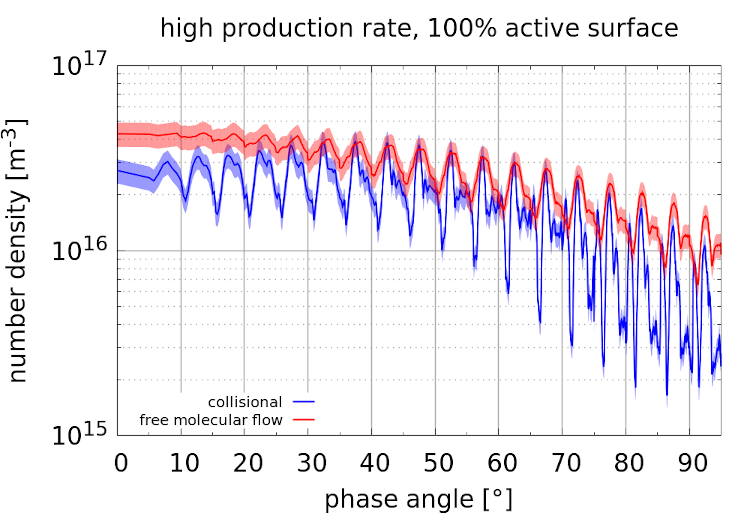}
  \caption{Local gas number density [m$^{-3}$] as a function of phase angle comparing the 100\% active surface models assuming the full physics of collisions (blue) and free molecular flow (red) for the high gas production rate.}
  \label{fig:rosina-highQ_free} 
\end{figure}

\newpage
% References
\bibliographystyle{aa}
\bibliography{references}

\end{document}